%% file: elsarticle-template-harv.tex
\documentclass[preprint,12pt,authoryear]{elsarticle}

\usepackage{amssymb}
\usepackage{amsmath}

\journal{International Journal of Human-Computer Studies}
\input{macro.tex}
\begin{document}

\begin{frontmatter}



\title{IntenBot: Flexible and Imprecise Multimodal Input for LLMs to Understand User Intentions for Casual and Human-Like HRI} 



            
            

\author[1]{Yen-Ting Liu}
\ead{owen4096@gmail.com}

\author[2]{Chiu-Hsuan Wang}
\ead{chwang821014@ntu.edu.tw}

\author[1]{TzuLing Chen}
\ead{chentzuling.nccu@gmail.com}

\author[2]{Ting-Ying Lee}
\ead{tyleeeeee815@gmail.com}

\author[3]{Tzu-Hua Wang}
\ead{wangtzuhua1002@gmail.com}

\author[3]{ChienMing Lin}
\ead{jianming1481@gmail.com}

\author[2]{Bing-Yu Chen}
\ead{robin@ntu.edu.tw}

\author[1]{Hsin-Ruey Tsai\corref{cor1}}
\ead{hsnuhrt@gmail.com}

\cortext[cor1]{Corresponding author}
\address[1]{National Chengchi University}
\address[2]{National Taiwan University}
\address[3]{CTO Office, Delta Electronics}






\begin{abstract}

In natural human-to-human communication, multimodal user input is typically used to supplement explicit and complement implicit voice commands, with casualness allowing for flexible input modality combinations and tolerance for imprecise input data.
For example, saying \textit{``I want that.''} with a casual glance at a bottle of water is clear enough in human-to-human communication as an implicit voice command accompanied by gaze and/or gestures, rather than an explicit one.
To enable such a human-like interaction in human-robot interaction (HRI), we propose a system, IntenBot, to understand user intentions from flexible and imprecise multimodal input, including voice, gaze, and finger-pointing, in XR.
The disambiguation capability of large language models (LLMs) is used to filter out irrelevant input modalities and imprecise input data, generating potential instructions for user confirmation. 
The flexible and imprecise multimodal input enables casual, human-like interaction with robots, reducing time, effort, and attention, and could also be used as non-voice input.
We conducted an informative user behavior study in a simulated environment to understand users' natural behavior in flexibly interacting with a robot using multimodal input and to obtain appropriate angle range parameters for gaze and finger-pointing.
An XR study was then performed to evaluate the performance of IntenBot, compared with other methods. 
We also deployed IntenBot on a physical robot to showcase its real-world applications.
\end{abstract}

\begin{graphicalabstract}
  \includegraphics[width=1\linewidth]{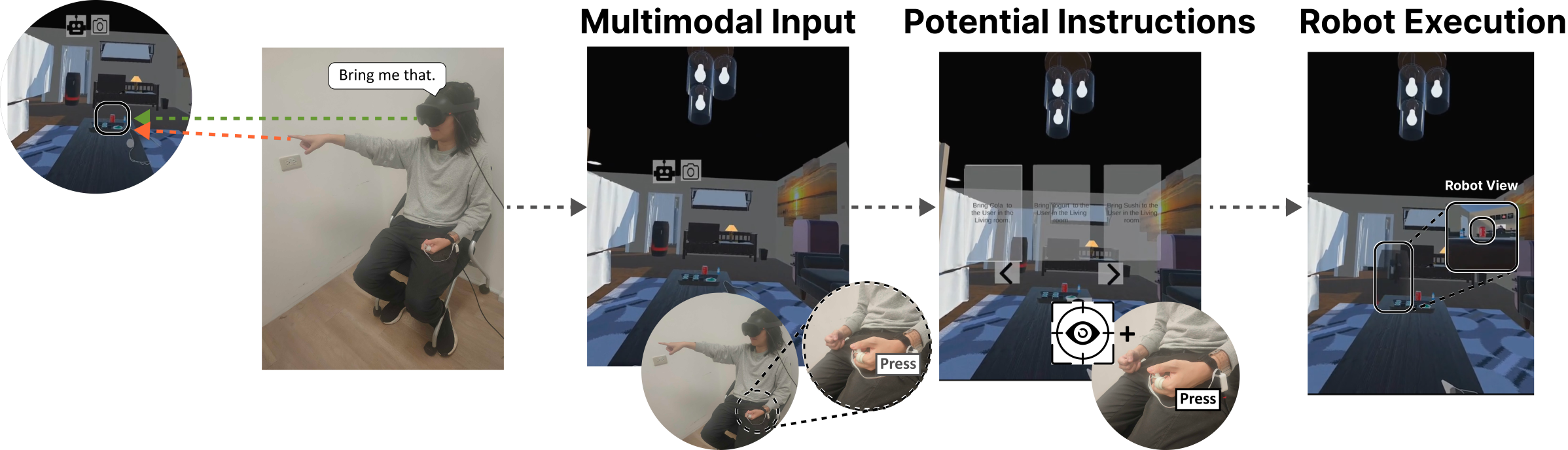}
\end{graphicalabstract}

\begin{highlights}
    \item Proposing an interaction paradigm and implementing a system to enable flexible and imprecise multimodal input for HRI, as in casual, natural behavior in human-to-human communication, using LLM disambiguation.

    \item Understanding casual, natural multimodal user behavior in HRI and obtaining appropriate angle range parameters for gaze and finger-pointing.

    \item Evaluating the system performance and verifying that it outperforms the other methods, especially in word count, mental demand, perceived efficiency, and the conversation condition, due to flexibility; also demonstrating its use as non-voice input and in real-world robot applications.

\end{highlights}


\end{frontmatter}



\begin{figure}[ht]
    \centering
    \includegraphics[scale=0.16]{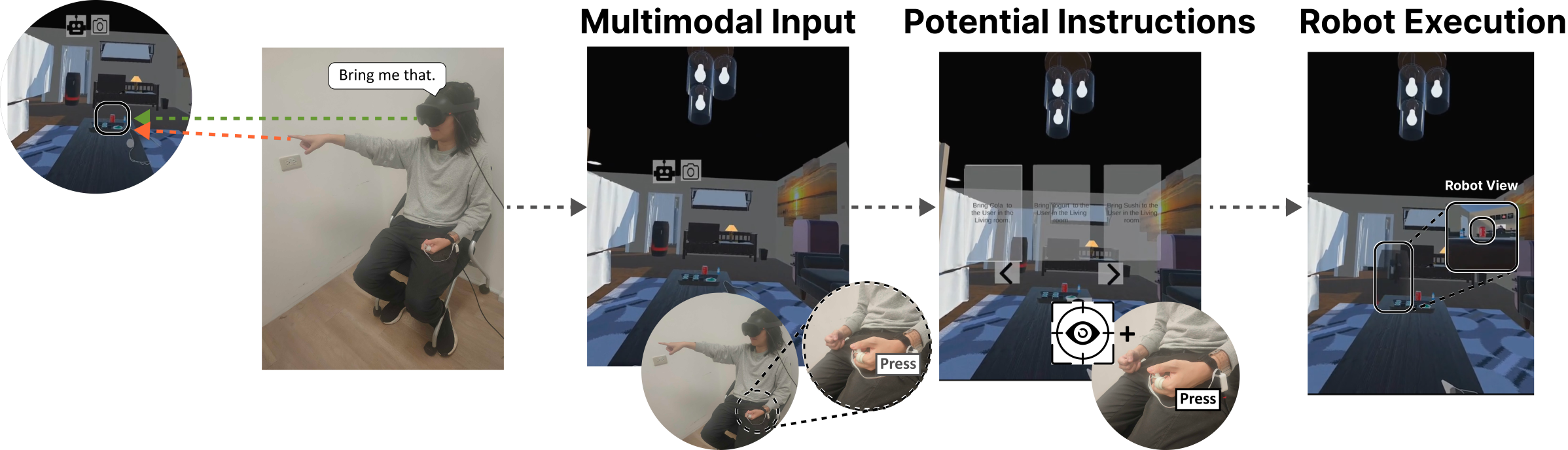}
    \caption{IntenBot allows users to use flexible and imprecise multimodal input for casual and human-like interaction with robots. 
   Users can use implicit voice input \textit{``Bring me that.''} combined with gaze and finger-pointing input to issue a command (left). 
   The thumb touches and presses a ring to activate the system and handle multimodal input timing (middle-left).
   IntenBot leverages LLMs to disambiguate flexible and imprecise multimodal input, interpret user intentions, and generate potential instructions for user confirmation (middle-right).
   The robot then executes the task of bringing the cola (right).
}
    \label{fig:teaser}
    \vspace*{-6pt}
\end{figure}

\section{Introduction}

Several input methods have been used in human-robot interaction (HRI), \eg voice command (\cite{ahn2022can,xiao2024robi,kim2024understanding}), gestures (\cite{quintero2013sepo,waldherr2000gesture,lin2023gestureinformed,bolt1980put}) and multimodal user input through robot multimodal sensors (\cite{wang2024lami,qin2023multimodal}).
However, there is still a gap between interacting with robots as one would with humans or even close friends who understand each other well, and interacting with them as passive entities. 
For instance, if a user says \textit{``I want that''} while roughly pointing at the can of juice on the table with a finger, it is clear enough for a friend to understand which object is indicated.
This works even if the rough finger-pointing may be imprecise and the eyes glance elsewhere, without needing to explicitly say \emph{``Please bring me the can of juice on the table in front of you.''}.
This shows that implicit voice commands accompanied by gaze, gestures, and/or body language, rather than explicit voice commands, are common in human-to-human communication to reduce time, effort, or even attention, especially when engaged in a primary task.
Higher familiarity with the space, leading to better spatial understanding, could further benefit implicit commands. 
Furthermore, the casualness of natural human-to-human communication leads to \textit{flexibility} in using various combinations of multimodal input methods and \textit{imprecision} in input data.
Therefore, multimodal user input with the ability to disambiguate irrelevant modalities and imprecise input data is critical for robots to interpret users' implicit commands and behavior to understand their intentions.

Multimodal user input has been used in HRI to enhance precision 
(\cite{perzanowski2001building,bolt1980put}). 
However, current systems require users to follow predefined and prescribed input methods, such as using predetermined gesture sets (\cite{iba2005interactive,lin2023gestureinformed,perzanowski2001building}) or precisely indicating targets through gaze or finger-pointing (\cite{quintero2013sepo,perzanowski2001building}), within specific modality combinations.
These constrain flexibility, demand precise input, and require users to learn and remain attentive when interacting with robots, since they do not filter out irrelevant modalities or imprecise data. 
For voice input in HRI, natural language processing (NLP) and large language model (LLM) techniques have been leveraged to parse users' commands in the context of their natural language and environments, thereby refining the instructions given to robots or voice assistants (VAs) (\cite{lin2023gestureinformed,lee2024gazepointar}). 
LLMs help interpret users' imprecise input and incorporate environmental and robotic data into planning. 
Although these systems allow flexible and imprecise voice input, they are limited to a single input modality.
LAMI (\cite{wang2024lami}) preliminarily showcases the potential of LLMs to disambiguate multimodal user input, including speech, gaze, pose, and actions, to understand users' intentions.
However, multimodal input systems with disambiguation capabilities to allow flexible modality combinations and handle imprecise input data have not been fully explored in current HRI.




We propose a system, IntenBot, to understand users' intentions from flexible and imprecise interactions with robots by leveraging multimodal input for LLMs (\figname\ref{fig:teaser}). 
Common modalities in human-to-human communication, including voice, gaze, and finger-pointing gesture input, are included in IntenBot. 
An extended reality (XR) headset is used to collect data, even when robots are not around users. 
The disambiguation capability of LLMs is leveraged to analyze the data and understand users' instructions, even with irrelevant input modalities or imprecise input data.
The potential instructions are then displayed for users to confirm.
The system achieves input flexibility and tolerance for imprecision to reduce users' time, effort, and attention.
It can even be used as a non-voice input method relying solely on non-voice modalities, allowing robots to be instructed in noisy or quiet scenarios, \eg factory or library.
We conducted an informative user behavior study to understand how users communicate with robots in a flexible human-to-human-like manner in a simulated environment, and to obtain appropriate angle range parameters for gaze and finger-pointing. 
Based on the results, an XR experience study was performed to evaluate the overall performance of IntenBot compared with other methods.
The IntenBot system was also deployed on a physical robot to showcase several applications.

The paper presents the following contributions:

\begin{enumerate}
    \item Proposing an interaction paradigm and implementing a system to enable flexible and imprecise multimodal input for HRI, as in casual, natural behavior in human-to-human communication, using LLM disambiguation.

    \item Understanding casual, natural multimodal user behavior in HRI and obtaining appropriate angle range parameters for gaze and finger-pointing.

    \item Evaluating the system performance and verifying that it outperforms the other methods, especially in word count, mental demand, perceived efficiency, and the conversation condition, due to flexibility; also demonstrating its use as non-voice input and in real-world robot applications.

\end{enumerate}


\section{Related Work}

\subsection{Multimodal User Input}

Compared to single-modality user input, such as voice or gaze input, multimodal user input combines multiple input forms to more effectively convey users' intentions.
For instance, integrating speech with gestures (\cite{iba2005interactive, mahadevan2021grip, perzanowski2001building, qin2023multimodal, xiao2024robi}), speech with gaze (\cite{roider2018just, khan2022integrating, miniotas2006speech}), and gestures and gaze (\cite{pfeuffer2014gaze,chatterjee2015gaze+, lystbaek2022gaze}) enhances input precision and expressiveness.
Furthermore, the non-voice gaze and gesture combinations provide an alternative when speaking is inconvenient. 
In addition to enhancing precision, multimodal input methods are also used to facilitate the interpretation and understanding of users' intentions. 
``Put-that-there'' (\cite{bolt1980put}) illustrates how pointing gestures can clarify semantic ambiguities.
\cite{ghamandi2024unlocking} investigated the effectiveness of various input modality combinations in VR remote collaboration. 
Furthermore, multimodal input has improved robots' comprehension of user instructions. 
\cite{perzanowski2001building} developed a system that enabled robots to understand and respond to combined speech and gesture commands. 
Similarly, 
\cite{qin2023multimodal} demonstrated the effectiveness of integrating gaze, hand gestures, muscle gestures, and touch input, allowing robots to better understand users' intentions and deliver more contextually appropriate responses. 
These works underscore the benefits of multimodal input in enhancing the interpretation of user instructions.
However, they still require users to follow predefined interaction protocols, inducing a learning curve and increasing their cognitive load.


\subsection{LLMs in Human-Robot Interaction}

The development of LLMs has revolutionized HRI through their capability and flexibility to process complex data and generate diverse task formats. 
Unlike traditional rule-based robotic planning, which limits adaptability, LLMs incorporate contextual understanding to generate flexible and relevant task plans. 
By analyzing environmental data and decomposing high-level objectives into actionable steps, robots can operate more autonomously in various contexts (\cite{rana2023sayplan, ahn2022can, silver2022pddl, wake2023chatgpt,vemprala2024chatgpt, joublin2024copal,yoneda2024statler,zha2024distilling}). 
Moreover, integrating function libraries allows LLMs to generate robot code directly, expanding their applicability in HRI (\cite{izzo2024btgenbot,ge2024cocobo,karli2024alchemist}).
In addition, LLMs can process natural and ambiguous data, enabling more advanced user interfaces. 
They enable a transition from predefined commands to more sophisticated, multi-turn dialogues (\cite{huang2022inner}). 
LLMs can also infer users' intentions even when instructions are ambiguous or incomplete (\cite{zhao2023erra}). 
EMMA (\cite{yang2024embodied}) can process users' textual instructions alongside environmental images to generate action sequences for efficient task execution. 
\cite{vemprala2024chatgpt} proposed a pipeline allowing non-technical users to provide high-level feedback to an LLM for monitoring robot performance.
\cite{kim2024understanding} further showed expectations for the potential of LLM-powered robotic systems to interpret non-verbal cues. 
These advancements highlight the potential of LLMs in HRI, particularly for processing complex and ambiguous data. 
However, previous works primarily focus on disambiguating users' voice and textual instructions with LLMs, even with environmental data, rather than leveraging this disambiguation capability to interpret and understand instructions from multimodal user input.

\subsection{User Intention Disambiguation}
To resolve ambiguous instructions, prior works have utilized multiple modalities, \eg visual, auditory, and contextual data, to clarify users' intensions (\cite{wong2010you,wong2014support,quintero2013sepo}). 
Machine learning techniques make these solutions more adaptable and capable of resolving ambiguities across various scenarios(\cite{mayer2020enhancing,goel2023you,kong2014you,kottur2021simmc,guo2022gravl}). 
This enables the system to interpret unclear input by leveraging data from multiple modalities, such as recognizing visual cues or contextual hints when verbal commands are ambiguous.
Recent advancements in LLMs have significantly improved approaches for understanding environmental information and clarifying ambiguous user instructions. 
Environmental contextual data, such as point clouds (\cite{hong20233d}) and images (\cite{rana2023sayplan}), can be interpreted and structured as scene graphs. 
Combining such environmental data with other user input, like voice commands or pointing gestures, enables more precise disambiguation and inference of user intentions (\cite{xiao2024robi,lin2023gestureinformed}).
GazePointAR (\cite{lee2024gazepointar}) utilizes gaze, pointing gestures, and conversation history to resolve speech query ambiguities via an augmented reality (AR) headset. 
LAMI (\cite{wang2024lami}) integrates human and environmental factors, including speech, gaze, posture, and object placement, to generate dynamic, multimodal expressive output for robotic systems.
However, current disambiguation methods focus on handling multimodal input with correct data.
Nevertheless, in natural human-to-human communication, casualness often introduces noise or missing data.
IntenBot focuses on addressing this by leveraging LLMs to provide greater input flexibility, tolerating irrelevant modalities and imprecise data.



\section{IntenBot}
We propose a system, IntenBot, that enables users to interact with a robot in a casual manner, as in natural human-to-human communication, allowing for flexible and imprecise multimodal input.
It leverages multimodal user input for LLMs to disambiguate various modality combinations and imprecise input data to infer users' intentions for both explicit and implicit commands in natural behavior. 
Thus, IntenBot enables users to engage in natural, unconstrained robot interactions without predefined interaction protocols.

\subsection{Design Considerations}
To accomplish the goals, two design considerations are critical.

\begin{itemize}
    \item \textit{User Input Versatility.}
    To support users’ natural behavior without constraints, the system must incorporate versatile user input modalities in common human-to-human interactions, such as voice, gaze, gesture, facial expression, and body language input.
    Including more modalities is essential for versatility.
    Furthermore, without predefined protocols, the choice of modality combinations can vary even for the same user, especially under different environmental, social, and privacy conditions, \eg when engaging in a primary task, with hands occupied, in noisy environments, or during a conversation. 
    Thus, flexibility in using various multimodal input combinations, tailored to users' preferences and situations, is another key requirement for versatility.

    \item \textit{Disambiguation.}
    To achieve multimodal input versatility and flexibility, the system must be capable of resolving ambiguous input combinations or imprecise input information, ensuring natural and unconstrained interaction experiences. 
    Without predefined input modalities, the system receives information from both relevant and irrelevant modalities.
    Therefore, the system must be able to filter out data from irrelevant input modalities.
    Furthermore, imprecise input data may result from casual behavior or differing modality relationships, such as gaze input supplementing explicit voice input, or finger-pointing input complementing implicit voice input with pronouns.
    Therefore, the disambiguation capability is essential for enabling robust and flexible multimodal input with tolerance for imprecision.
    Accordingly, LLMs are well-suited to these requirements.
\end{itemize}

\begin{figure}[ht]
    \centering
    \includegraphics[scale=0.09]{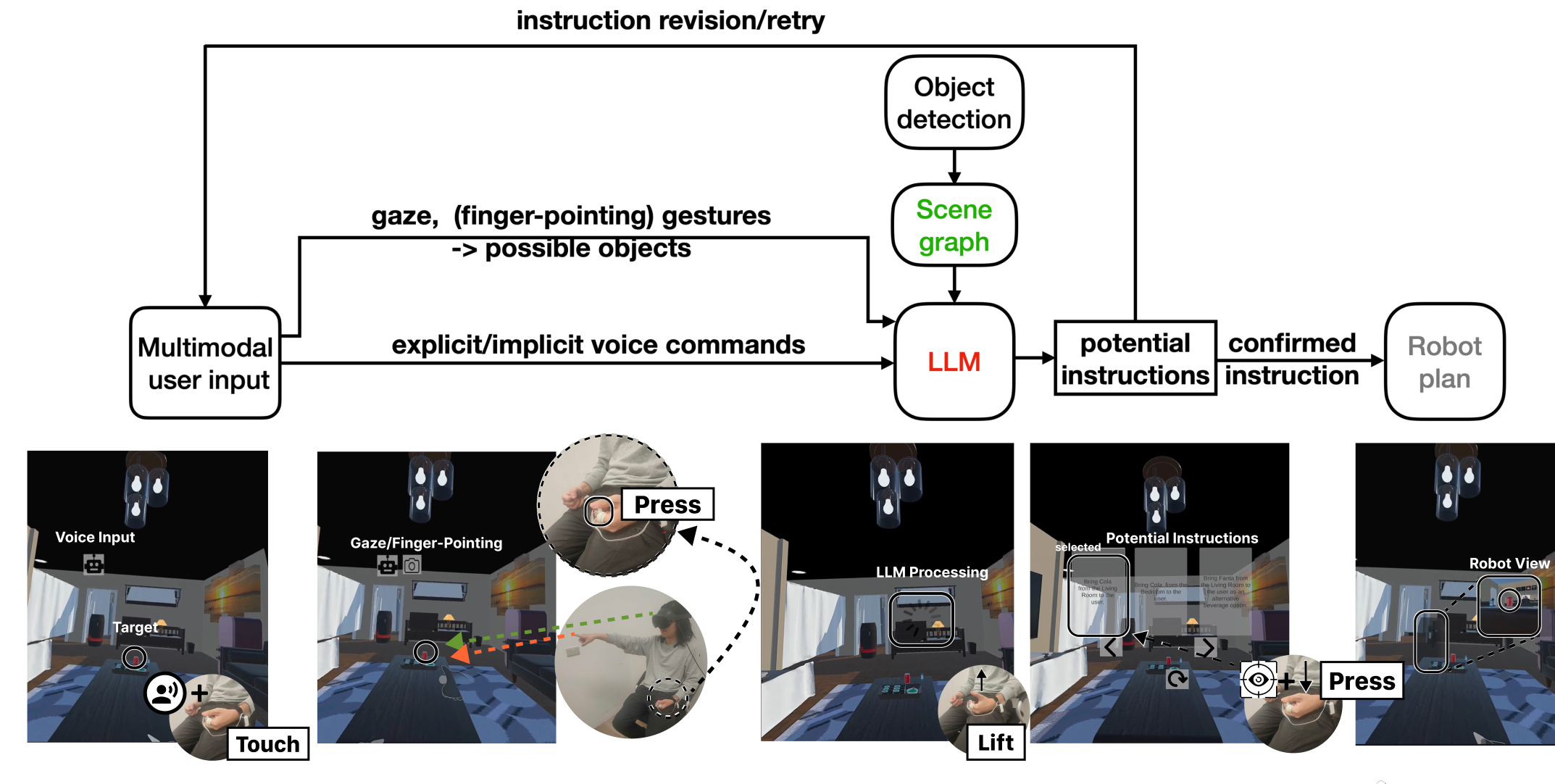}
    \caption{The flow of the IntenBot system.}
    \label{fig:SystemFlow}
    \vspace*{-6pt}
\end{figure}

\subsection{Implementation}
To achieve the design considerations, we propose IntenBot, which leverages an XR headset to collect multimodal user input data and utilizes the disambiguation capability of the LLM technique for intention inference and command interpretation. 
The IntenBot system flow is shown in~\figname~\ref{fig:SystemFlow}.
The system first collects multimodal input via the XR headset, uses the data as input to the LLM to generate \textit{potential instructions}, and then allows the user to confirm the correct one.
This design enables users to interact with a robot using flexible and imprecise multimodal input, as in their casual, natural behavior in human-to-human communication.
The confirmed instruction is further used to generate ROS 2 code for robot execution.
The main contribution of this work is the proposed interaction paradigm enabled by IntenBot, so details of the robot plan and control are in Section~\ref{sec:generalizability} and~\figname~\ref{fig:robot}, where a robot controlled by the IntenBot system is demonstrated.
The complete system, including details of the robot plan, is shown in~\figname~\ref{fig:wholeflow} in the appendix.

The multimodal input in IntenBot includes voice command input, gaze input, and finger-pointing gesture input, all captured by a Meta Quest Pro XR headset. 
We envision future XR headsets becoming smaller and lighter, making them suitable media for daily use and enabling multimodal input collection even in remote robot operation. 
Voice commands are captured by the headset's built-in microphone and transcribed using the Whisper API.
Gaze and hand gestures are tracked by the headset, while finger-pointing gestures are recognized using the Meta XR SDK, which determines the extended finger(s) and their pointing direction(s). 
Notably, since the two thumbs and two index fingers are mainly used for pointing, the direction(s) are recorded only for these four fingers when extended.
Moreover, no visual feedback is provided for the gaze point or the finger-pointing raycast, making it more compatible with various XR headsets and preventing interference with real-world interactions.

To activate the system, handle multimodal input timing, and confirm instructions, a ring with a button and an infrared (IR) sensor on the button, as in GazeNoter (\cite{tsai2025gazenoter}), is used. 
This setup provides touch and press input, even in a touch-and-press step.
The ring is worn on the non-dominant index finger, and the button can be positioned in the thumb's comfort zone (\cite{tsai2016thumbring}), allowing easy touch and press.
Users touch the button to activate the system and start recording a voice command, while a robot icon is displayed on the headset.
Without lifting the thumb, they can press the button with click feedback to capture the gaze and finger-pointing directions at that moment, while a camera icon appears alongside the robot icon.
When the thumb lifts off the button, the voice recording ends, and the multimodal command is sent to the LLM for processing, with a loading icon shown.
The design allows users to perform voice commands and the other input modalities (\ie gaze and finger-pointing input) simultaneously, with distinct timing for each modality.
Notably, to point at multiple objects in sequence, users can press the button multiple times without lifting the thumb.


The input to the LLM (GPT-4o) includes transcripts of explicit and/or implicit voice command(s) and \textit{possible object(s)}, which are determined from the gaze and finger-pointing directions and represent what users may be indicating.
Possible objects are derived from up to five pointing directions: gaze, two thumbs, and two index fingers, if these four fingers are recognized as extended.
Possible objects based on gaze are always detectable since gaze is continuously tracked.
The data is processed by the LLM to generate potential instructions. 
Its prompt includes five key components: guidance, rules, input data, output format, and skill library.
The guidance introduces the LLM to the overall task, instructing it to suggest potential user-intended actions from the input data. 
The rules guide decision-making by preventing the generation of unreasonable or repetitive tasks, ensuring proper use of external functions from the skill library when needed (\eg for mathematical calculations), and considering user behavior, such as gaze or finger-pointing, as indicators of intentions.
Furthermore, to reduce ambiguity in interpreting multimodal input, a modality priority policy is included: voice commands have the highest priority, while gaze and finger-pointing are treated equally. 
The input data includes the user’s current position and the formatted multimodal input, helping the model filter out irrelevant input and infer the user’s intention. 
The output format instructs the model to generate both the task content shown to the user and an explanation, ensuring that the suggestion is reasonable and traceable.
The skill library provides the model access to external computational tools, such as distance calculation functions, enabling more precise and informed decision-making.
The prompt is in~\figname~\ref{fig:prompt1} and~\ref{fig:prompt2} in the appendix.

The LLM also receives environmental data as a scene graph with node-based representations.
These encode object identity, spatial relations, and semantic affordances to help identify target objects for potential interactions. 
To capture and reconstruct the physical environment, an RGB-D camera (iPad Pro) with the Record3D application is used to record 3D spatial data. 
This data is then processed using the ConceptGraph model (\cite{gu2024conceptgraphs}) to generate a structured scene graph, which is used in the demonstration in Section~\ref{sec:generalizability}. 
In our user studies, which used a simulated environment, the scene graph data was manually created by the authors to match the virtual setup.
This environmental data, from either physical or simulated environments, also includes information about objects outside the user's field of view (FoV) or pointing range.
This allows the system to tolerate imprecise interactions and provides the model with a broader scene context, enabling it to make informed decisions even without direct pointing or gaze input.
Besides, we intended to include unpredefined gestures beyond deictic pointing gestures, recognized by a visual language model (VLM) (GPT-4o), as an auxiliary input in a pilot study described in the appendix and~\figname~\ref{fig:vlm}. 
However, the slow response time (over 40 seconds) prevents its use in our real-time system.

After the disambiguation process, the LLM interprets the user's command and infers their intention by generating 9 \textit{potential instructions}, ranked from most to least likely.
3 of them are shown on the headset at the same time, so the user can review all by switching at most twice.
The user selects the correct instruction to confirm, which is sent to the robot plan, described in Section~\ref{sec:generalizability}. 
If none are correct, the user can revise or retry via the retry icon.
All selection uses gaze selection and ring confirmation.
The IntenBot system takes about 15 seconds to generate the potential instructions via the LLM after receiving multimodal user input. 
Through the interaction design, IntenBot enables flexible and imprecise multimodal input, as in casual, natural human-to-human communication.

\section{Informative User Behavior Study}
To investigate casual, natural user behavior when employing multimodal input to interact with a robot, we conducted an informative user behavior study. 
Their behavior and performance were analyzed across conditions to provide insights and guidelines for our system.
Through this study, we also obtained appropriate angle range parameters for gaze and finger-pointing, reflecting casual, natural behavior in human-to-human communication.
Due to flexible and imprecise input with LLM disambiguation, these parameters differ from those in previous object or item selection research. 
The results are also used to complete our IntenBot system.
Notably, to observe natural user behavior, the system collected data without visual feedback, so users were unaware of whether their commands to the robot were interpreted correctly.
Performance evaluation and comparison were conducted afterward using the collected data.

\subsection{Apparatus and Participants}
An Oculus Quest Pro headset and a ring (five sizes, 18-22 mm diameter) were worn. 
We recruited 12 participants (7 female), aged 22-28 years (mean = 24.3).
All were non-native English speakers, and 10 of them were right-handed. 
Each received a \$10 compensation for their time and participation.

\begin{figure}[ht]
    \centering
    \includegraphics[scale=0.25]{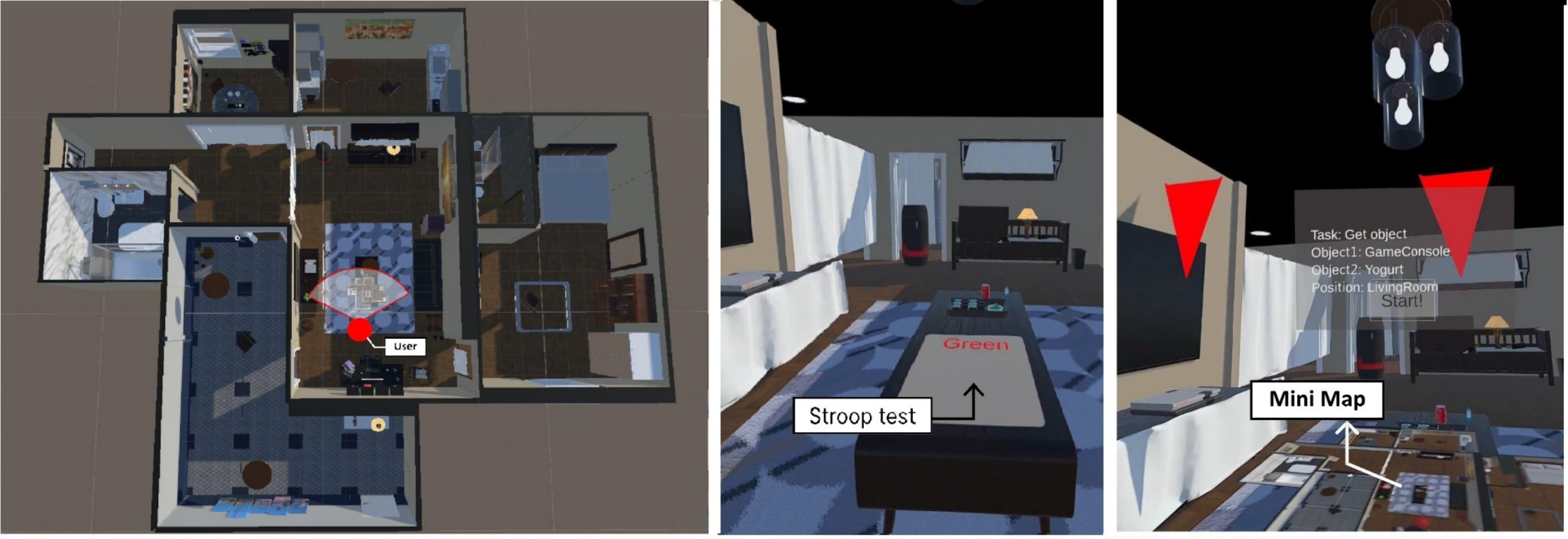}
    \caption{The VR home-like scene for the user behavior study (left). The Stroop test condition (middle). The task reminder and the mini-map with the marked target object(s) (right).}
    \label{fig:VRsetup}
    \vspace*{-6pt}
\end{figure}

\subsection{Task and Procedure}

A VR scene with 7 rooms and 95 objects was built to simulate a home-like setting (\figname~\ref{fig:VRsetup}), where participants instructed a virtual robot to bring object(s) to them or move an object to a target destination.
Since a large space with many objects, like a large apartment, was required to explore different conditions in this study, it was difficult to examine in the real world. 
Thus, the simulated VR scene was used, which is a common and acceptable study method in HRI (\cite{mahadevan2021grip, helgert2024unlocking}).
Three interaction methods: \emph{voice-only (explicit)} (V), \emph{multimodal input (explicit)} (E) and \emph{multimodal input (implicit)} (I), were examined as the primary independent variable.
\emph{Voice-only (explicit)}, the only single-modal input method, required complete verbal commands with all necessary details, as in common VAs, including the task, target object(s), and their location(s) or also destination, \eg \textit{``Please bring me the wine in the living room.''} 
\emph{Multimodal input (explicit)} also required fully detailed voice commands but encouraged supplementing them with other modalities, including gaze and/or finger-pointing, to further clarify intentions. 
In contrast, \emph{multimodal input (implicit)} asked for voice commands that intentionally omitted at least one key detail, to be inferred from other modalities, \eg the command that \textit{``Bring me that''} accompanied by a finger pointing toward the wine bottle. 
By comparing (V) and (E), we could understand whether multimodal input assists explicit voice commands in enhancing accuracy.
By comparing (I) to the others, we could observe whether implicit voice commands with multimodal input could reduce input time and word count while maintaining accuracy.
All methods used the IntenBot system to collect input from various modalities for data collection, without visual feedback.

Furthermore, this study included three additional factors as independent variables, including task horizon type, object visibility, and occupation condition, to observe natural behavior across various conditions and examine whether specific methods were preferred in certain conditions.
Two task horizon types common in HRI (\cite{zhao2023erra}) were included: short-horizon tasks involving single-step planning and long-horizon tasks involving multi-step planning. 
Short-horizon tasks required fetching a single object, \eg bringing the wine from the living room, while long-horizon tasks involved fetching multiple objects, \eg bringing the wine and apple juice from the living room, or moving an object to another location, \eg moving the wine to the dining room. 
This factor allowed us to observe participants' behavior with both simple and complex commands and evaluate the system's understanding of varying complexity.
Three target object visibility conditions were tested: clearly visible in the same room, hidden within another object or furniture in the same room, and in a different room.
This factor allowed us to observe how visibility influenced participants’ multimodal input strategies.
Three occupation conditions were examined: non-occupied, in a Stroop test requiring visual attention, and in a conversation.
The Stroop test condition simulated participants being occupied with a primary task (\eg working or reading). 
A color word (\eg ``yellow'') was shown in a random font color (\eg green), with both text and font color chosen from blue, red, green, and yellow, and refreshed every 2.5 seconds (\cite{tsai2021guideband, stroop1935studies}). 
They counted how many times the word meaning matched the font color, maintaining at least 75\% accuracy.
In the conversation condition, they conversed with an avatar controlled by the experimenter, requiring attention and increasing mental pressure due to social acceptance (\cite{tsai2025gazenoter}).
This factor allowed us to observe participants' behavior and strategies across various occupation conditions.

At the beginning, participants were introduced to the experiment and system, calibrated eye tracking, sat on a chair matching the VR living room setup, and familiarized themselves with the system and VR environment in a training session. 
In the study, each task trial consisted of a preparation phase and a task phase.
In the preparation phase, a task reminder was shown, and participants organized the command and strategy to issue to the robot based on the examining method and conditions. 
They could freely move around in the VR scene to examine the target object(s) and become familiar with their positions and/or destination.
A mini-map with the marked target object(s) and participants' current position helped them build spatial understanding.
Since implicit multimodal input requires environmental familiarity, this ensured that all participants had a similar level of familiarity with the VR scene, despite the prior training session. 
They pressed the button on the ring to enter the task phase and were returned to the initial chair position.
A ``Ready'' reminder appeared, followed by ``Start'' after a random 3-7 second delay, which occupied participants during the Stroop test and conversation conditions. 
They then issued the command using the examining method. 
The trial ended when they delivered the command by lifting their finger off the ring.
The LLM processing, instruction display, and selection in IntenBot were excluded from this study.
The experimenter ensured that the command matched the examining method and conditions.

A total of 54 (= 3 (methods) $\times$ 2 (horizon types) $\times$ 3 (visibility conditions) $\times$ 3 (occupation conditions)) trials were examined, with counterbalanced method order and randomly assigned horizon types, visibility, and occupation conditions, ensuring even distribution.
A 10-minute break was between methods.
After the experiment, a semi-structured interview collected subjective feedback.
The study lasted about two hours. 
We collected various behavioral data, including task completion time (from ring touch to release to issue the command), audio recordings and verbal transcripts (also for word count), gaze and finger-pointing directions, and object information. 
Using these data, we further examined various gaze and finger-pointing angle ranges to determine selected objects, then combined them with the verbal transcripts as LLM input to evaluate the accuracy and obtain appropriate angle range parameters after the study.
The system was considered accurate if the ground truth task matched any of the 9 potential instructions generated by IntenBot, allowing users to select the correct one in real use.

\begin{figure}[ht]
    \centering
    \includegraphics[scale=0.102]{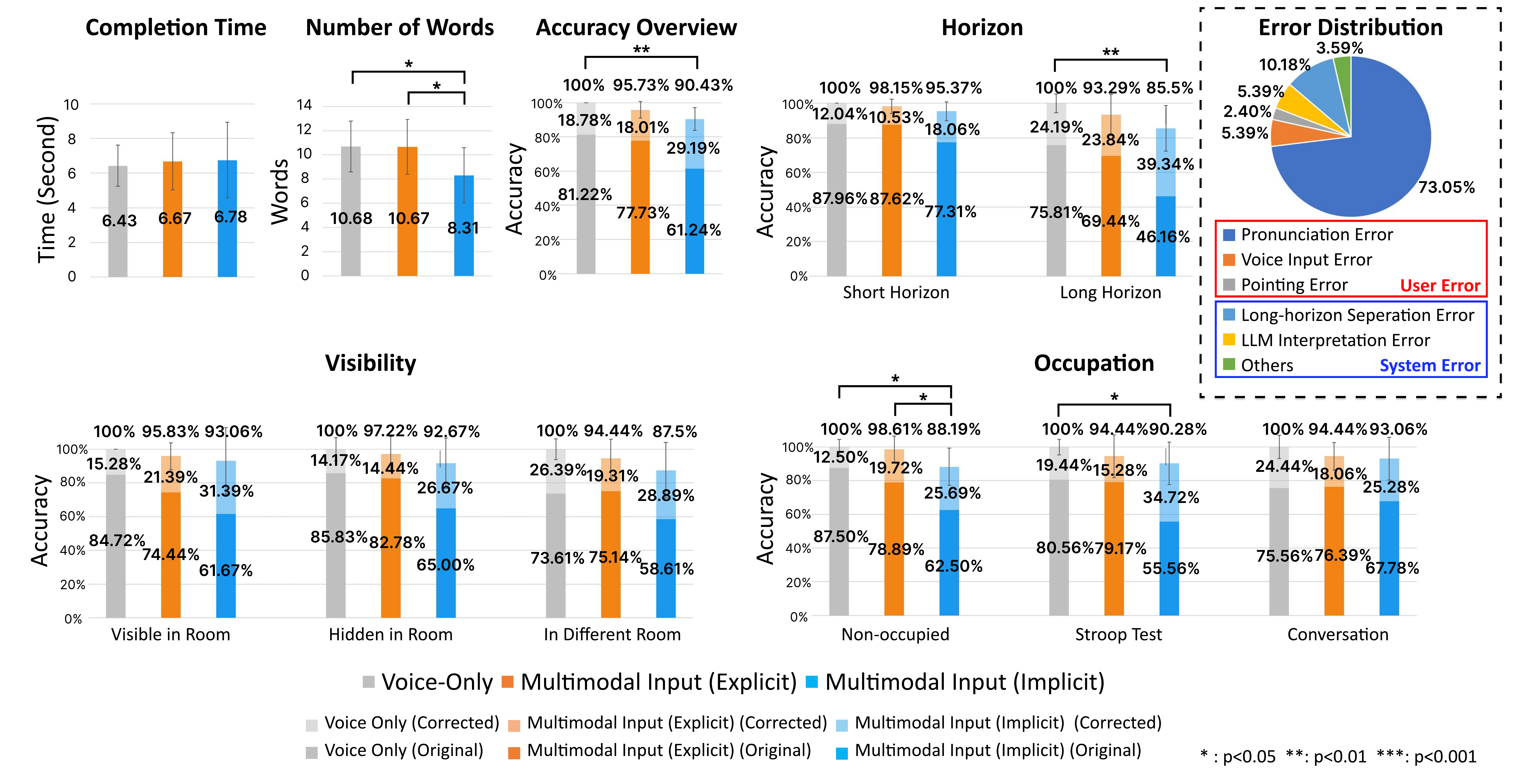}
    \caption{The study results of the informative user behavior study.}
    \label{fig:study1}
    \vspace*{-6pt}
\end{figure}

\subsection{Results and Discussion}

Repeated measures ANOVA and Bonferroni correction for post-hoc pairwise tests were utilized for the statistical analysis.
The results for are shown in~\figname~\ref{fig:study1}. 
No significant main effect ($F_{1.36, 14.96} = 0.31, p = 0.65$) was found across the methods for time, 
but a significant main effect ($F_{2, 22} = 9.31, p < 0.01$) was observed for the number of words per command.
The post-hoc pairwise tests revealed that the number of words in (I) was significantly lower than in (V) and (E), as expected.
For accuracy, we observed a severe pronunciation issue in the study due to non-native English speakers. 
Although all participants could fluently issue commands, the system might not have accurately converted the speech to text.
Since pronunciation was not the focus of this system, we manually corrected pronunciation errors using audio recordings, as in~\cite{kuhn2024measuring, wei2022could}, and then used the corrected voice command as LLM input to prevent them from influencing the evaluation, with original results also shown.
This is like using an advanced voice recognizer for diverse accents, without changing any words, meaning, and even other errors (including voice input errors or LLM interpretation errors), which are discussed later.

\begin{figure}[ht]
    \centering
    \includegraphics[scale=0.1]{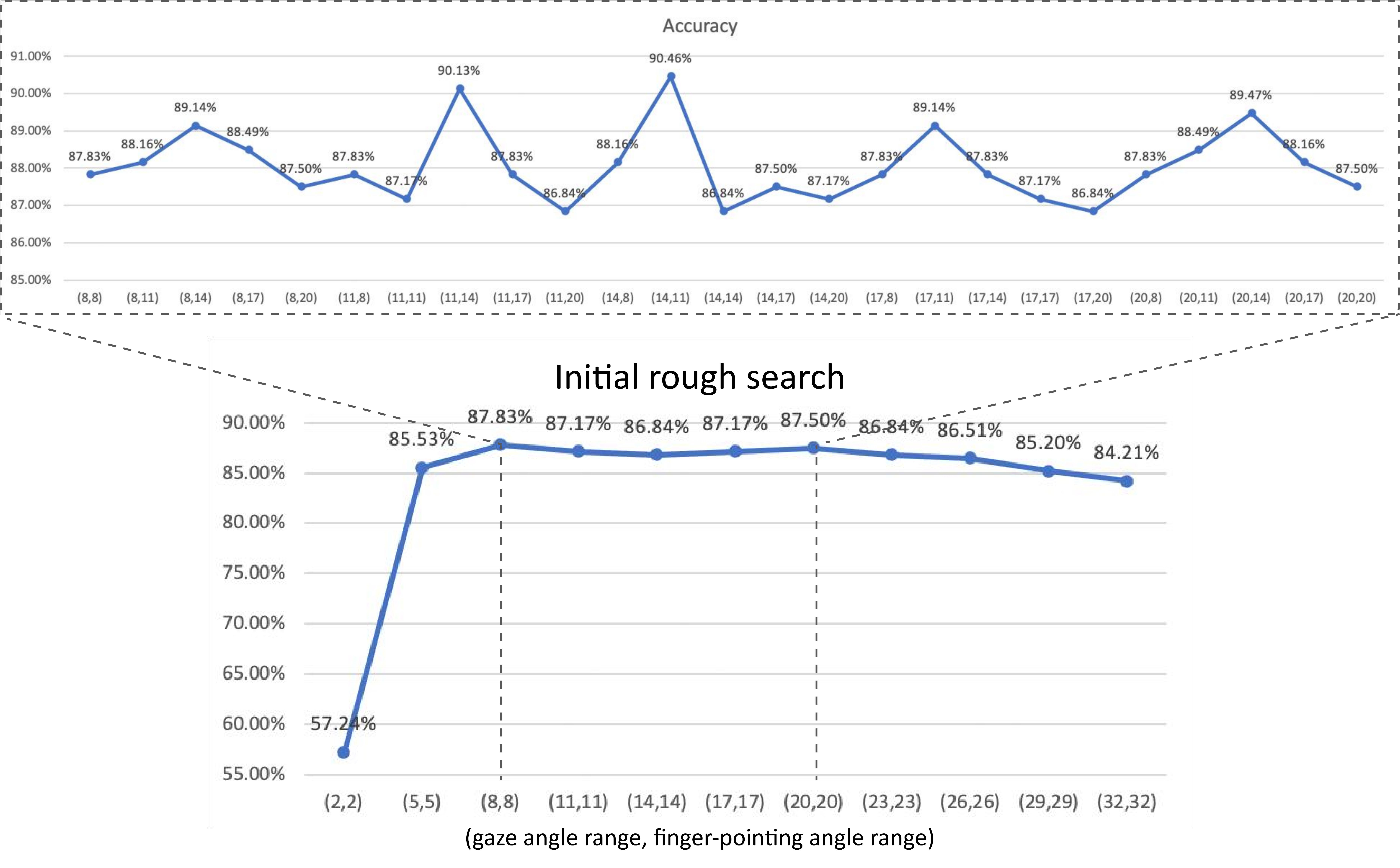}
    \caption{We began by evaluating the accuracy of various combinations of gaze and finger-pointing angle ranges to identify the optimal pair for use in the (E) and (I) interaction methods. In the initial rough search (Bottom), we tested angle ranges from $2^{\circ}$ to $32^{\circ}$, with $3^{\circ}$ increments for both gaze and pointing simultaneously. In the refined analysis (Top), we examined specific combinations of (gaze angle range, finger-pointing angle range) within these intervals, again using a $3^{\circ}$ step size. Finally, two peaks at ($11^{\circ}$, $14^{\circ}$) (90.13\%) and ($14^{\circ}$, $11^{\circ}$)(90.46\%) are identified.}
    \label{fig:Angle}
    \vspace*{-6pt}
\end{figure}

Before statistical analysis, we evaluated the accuracy across various combinations of gaze and finger-pointing angle ranges to obtain the combination with the highest accuracy for (E) and (I) (\figname~\ref{fig:Angle}).
Some trials in (E) were excluded to reduce noise, since participants did not use gaze or finger-pointing input to supplement the explicit voice command in their natural behavior.
Angle ranges, defined as angle distances from each raycasting direction, were initially examined from $2^{\circ}$ to $32^{\circ}$ in $3^{\circ}$ intervals as a rough search.
Higher accuracies occurred at $8^{\circ}$ and $20^{\circ}$, leading to a finer search, 
revealing two peaks at ($11^{\circ}$, $14^{\circ}$) (90.13\%) and ($14^{\circ}$, $11^{\circ}$) (90.46\%) in the combinations of \textit{(gaze angle range, finger-pointing angle range)}.
We finally selected $14^{\circ}$ for gaze and $11^{\circ}$ for finger-pointing, achieving the highest accuracy, as the appropriate parameters for the following accuracy analysis.
The accuracy of all trials across all three methods was 95.42\% after correction (originally 73.62\%),
with the error distribution in~\figname~\ref{fig:study1} (upper-right) indicating the original 26.38\% error rate. 
Pronunciation errors, the most severe at 73.1\%, 
were corrected and labeled as ``corrected'' data. 
Voice input errors (5.4\%) 
resulted from 
distractions, 
and pointing errors (2.4\%) occurred 
when the target(s) 
omitted in implicit voice commands fell outside 
gaze and finger-pointing ranges (only in (E) and (I)). 
These three error types originated from participants' behavior.
In contrast, system errors included LLM interpretation errors (5.4\%), caused by failure to disambiguate the target(s) and generate the correct instruction in the 9 potential instructions;
long-horizon separation errors (10.2\%), where multiple targets were mistakenly split into separate instructions;
and the other errors (3.6\%) involving various less frequent issues. 
Notably, long-horizon separation errors are not true errors but rather output format issues; they would be correct if users could select multiple instructions.
Therefore, they were also corrected and labeled as ``corrected'' data.
Among these, the LLM interpretation error is a key focus in this evaluation, showing its disambiguation capability with sufficient input information. 

For accuracy, only corrected pronunciation data were used in the statistical analysis.
A significant main effect was observed among the methods ($F_{2, 24} = 15.27, p < 0.001$) and for horizon types ($F_{1, 12} = 9.08, p < 0.05$), but not for visibility ($F_{2, 24} = 0.84, p = 0.52$) or occupation ($F_{2, 24} = 0.14, p = 0.74$).
The post-hoc pairwise tests showed that (V) had significantly higher accuracy than (I).
For horizon types, 
(V) had significantly higher accuracy than (I) in long-horizon tasks.
No statistical significances were found among visibility conditions.
For occupation, significant differences were revealed in the (V, I) and (E, I) pairs in the non-occupied condition, and in (V, I) in the Stroop test condition. 
The results indicate a trade-off between explicit verbal commands and implicit multimodal input. 
Although (I) required fewer spoken words, its task completion time was similar to (V) and (E). 
\textit{P4}, \textit{P5}, \textit{P6}, \textit{P8}, \textit{P9}, and \textit{P11} reported performing gaze and/or finger-pointing input after, not during, voice input. 
\textit{P4} specified that they were used to saying \textit{``Bring me that''} before finger-pointing.
This sequential input 
could explain why the completion time in (I) was not shorter as expected. 
Furthermore, although multimodal input supports intention understanding, flexible and imprecise input from casual, natural behavior led to inaccurate possible objects for the LLM, reducing disambiguation performance, 
especially in (I) lacking explicit voice commands. 
(E) and (I) still achieved 
over 95\% accuracy in simple tasks, \ie short-horizon tasks, offering flexible and tolerant input. 
However, (V) performed better in complex long-horizon and non-occupied tasks. 
Notably, this might be partly due to the task reminders in the preparation phase, which made it easier to issue explicit commands with little organization effort, as noted by \textit{P7}, \textit{P9}, and \textit{P12}, despite being part of the standard study design.


Participants' behavior was influenced by conditions. 
\textit{P2}, \textit{P4}, and \textit{P8} reported increased cognitive load in long-horizon tasks that required multi-step planning and simultaneous multimodal input in (E) and (I), especially under occupied conditions. 
The lack of voice content in (I) further reduced confidence and increased cognitive load.
For visibility, participants adjusted their strategies based on object location. 
8 participants (\textit{P1}-\textit{P5}, \textit{P7}, \textit{P10}, and \textit{P11}) used finger-pointing to supplement explicit commands in (E) and complement implicit ones in (I) for target(s) within their FoV, including hidden ones, in the same room.
Despite no significant improvement in completion time, only in word count, they still found multimodal input in (E) and (I) intuitive and helpful. 
5 of them (\textit{P2}-\textit{P4}, \textit{P7}, and \textit{P10}) even preferred combining gaze and/or pointing with voice input in all conditions in (E) and (I) to increase confidence in the spatial interaction of specifying object locations.
In contrast, when the object(s) were in other rooms, \textit{P1}, \textit{P5}, and \textit{P12} preferred detailed voice commands in (E) 
to maintain accuracy.
For occupation, all participants found explicit methods (V) and (E) 
challenging in the conversation condition due to social norms. 
Most waited for pauses before issuing voice commands, while \textit{P5} and \textit{P9} interrupted. 
In (I), lower verbal effort made 7 participants (\textit{P1}, \textit{P4}-\textit{P6}, \textit{P9}, \textit{P11}, and \textit{P12}) comfortable interrupting, supporting our assumption that (I) suits this condition and achieves accuracy comparable to (V) and (E).
In the Stroop test condition, high mental effort 
hindered gaze and finger-pointing in (E) and (I),
contrary to the expectation that finger-pointing would remain available. 

In general, results and comments show that (I) requires fewer voice input words but similar completion time as (V) and (E), due to inefficient sequential input habits.
Accordingly, a more efficient interaction design was explored in the following study.
Due to the low verbal effort, (I) is suitable during a conversation.
Furthermore, the obtained parameters ($14^{\circ}$, $11^{\circ}$) for \textit{(gaze angle range, finger-pointing angle range)} enable flexible and imprecise multimodal input in causal, natural behavior,
making (E) and (I), both advantages of IntenBot, perform well in short-horizon tasks (over 95\% accuracy) and preferred for spatial interactions, especially within the FoV.
Besides, LLM interpretation errors at 5.4\% validate its disambiguation capability and support the practicality of the interaction paradigm.

\section{XR Experience Study}

Based on the parameters and insights from the previous study, we finalized IntenBot and conducted this study to evaluate its overall performance, 
including the system response time, potential instruction feedback, and instruction revision. 
Furthermore, 
we explored a more efficient IntenBot interaction design method with brief training and 
the use of IntenBot as a non-voice input in this study.

The apparatus was the same as in the previous study.
We recruited 12 participants (2 female, 11 right-handed), aged 23-28 years (mean = 24.33).
None of them had taken part in the previous study.
All of them were non-native English speakers.
Each participant was compensated \$10 for their time and involvement in the study.

\subsection{Task and Procedure}
The same VR scene from the previous study was used.
Three methods were examined in this study: \emph{voice-only (explicit)} (V), \emph{IntenBot (free)} (F), and \emph{IntenBot (efficient)} (E). 
(V) was the previous \emph{voice-only (explicit)} method. 
(F) allowed natural, unconstrained multimodal input, similar to freely combining the previous \emph{multimodal input (explicit)} and \textit{multimodal input (implicit)} methods.
It represents the flexible, imprecise multimodal input proposed in IntenBot, as in casual, natural human-to-human communication.
(E), newly added, involved brief training for efficient IntenBot use, requiring simultaneous implicit voice commands and other modalities to reduce overall command time. 

Two strategies were taught in training of (E) to balance conciseness and clarity. \textit{Case 1}:
For short-horizon tasks with high visibility, participants said only the action verb (\eg \textit{``Bring''}, \textit{``Take''}, or \textit{``Get''}) and simultaneously pressed the button to indicate the target using gaze and/or finger-pointing.
\textit{Case 2}:
For long-horizon tasks and/or low visibility, participants said the target object name(s) while simultaneously indicating them with gaze and/or finger-pointing in fetching tasks; in moving tasks, said the target object name while indicating it, then the action verb (\eg \textit{``Move''} or \textit{``Go''}) when indicating the destination.
In case 1, with fewer possible objects as LLM input, the system could easily disambiguate the target. 
In case 2, high complexity in long-horizon tasks and/or low visibility made the target object(s) unclear, requiring naming the object(s) and/or action verb while using gaze and/or finger-pointing to disambiguate the locations and destination.
The strategies align with natural human-to-human communication, more information in complex or ambiguous conditions, and less in simpler, clearer ones.
Four examples were shown to illustrate the strategies, with each voice command containing one action verb, one or two object names, or one object name and one action verb.
Examples included: 
    (1) Saying \textit{``Bring''} while pressing the button and indicating the bottle of water with gazing and/or finger-pointing in the same room (Case 1: short-horizon, high visibility);
    (2) Saying \textit{``Whiskey''} while indicating it in another room. (Case 2: short-horizon, low visibility);
    (3) Saying \textit{``Sculpture''} and \textit{``Speaker''} while indicating each target (Case 2: long-horizon);
    (4) Saying \textit{``Pepper shaker''} while indicating it, then \textit{``Move''} while indicating the destination (Case 2: long-horizon).

Furthermore, \emph{IntenBot (non-voice)} (N), using only gaze and/or finger-pointing without voice input,
was included for additional comparison.
As a more restricted and less precise method, it was only examined in short-horizon tasks based on a pilot study.
Besides, although ($14^{\circ}$, $11^{\circ}$) for \textit{(gaze angle range, finger-pointing angle range)} were obtained by the previous study, we observed that too many possible objects within these ranges reduced accuracy, based on a pilot study.
We thus applied a priority policy for the LLM: possible objects within ($2.8^{\circ}$, $8^{\circ}$) were prioritized for precise selection, while those beyond but within ($14^{\circ}$, $11^{\circ}$) served as lower-priority, casual selections to enhance the system accuracy.
$2.8^{\circ}$ for gaze and $8^{\circ}$ for finger-pointing based on previous research
(\cite{kyto2018pinpointing, mayer2018effect}) differ from our $14^{\circ}$ and $11^{\circ}$, highlighting the need to determine the appropriate parameters in the previous study.
Furthermore, we improved the prompt to eliminate long-horizon separation errors in this finalized IntenBot system, with the prompt in the appendix.
By comparing (V) with others, we could observe whether IntenBot in various modes outperformed the baseline voice input. 
The (F, E) comparison revealed differences between natural behavior and efficient strategies, while (N) versus (F)/(E) verified IntenBot's practicality as non-voice input and its comparability to the other modes.
As in the previous study, three factors were included: task horizon type (short/long; (N) only in short), object visibility (same/different room), and occupation condition (non-occupied/in conversation). 
The conditions were simplified, since 
users showed behavior differences only in these two visibility conditions despite no statistical significance in the previous study, and
the Stroop test condition with high mental effort was found unsuitable for multimodal input.
The procedure followed the previous one but used the complete IntenBot system. 
Notably, the preparation phase included a task reminder and a see-through VR feature to mitigate bias toward voice input, especially in (V), as discussed in the previous results, 
by allowing participants to see through walls and objects to locate the target object(s) from the initial position, ensuring comparable familiarity and spatial understanding.
To prevent this feature from benefiting gaze and finger-pointing, they were asked to look down and lower their hands before the task phase. 
To address the pronunciation issue, they used their native language in this complete IntenBot system evaluation.
The system translated transcripts into English for the LLM, while the system text remained in English.
After issuing a command, they waited for the LLM to generate 9 potential instructions and selected the correct one.
If none were correct, they could revise and retry the task, up to two times. 
The first retry (second attempt) used the same method to allow behavior correction; the second (third attempt) allowed any method, since they might lose their patience and prefer a more precise method, except for (N), which remained restricted to non-voice input.
Trial were marked as correct if the correct instruction was selected within three attempts, including the two retries; otherwise, they were marked as errors.
The virtual robot then performed the task to conclude the trial. 

A total of 28 trials (= 3 (methods) $\times$ 2 (horizon types)  $\times$ 2 (visibility conditions) $\times$ 2 (occupation conditions) + 1 (additional method (N)) $\times$ 1 (horizon type)  $\times$ 2 (visibility conditions) $\times$ 2 (occupation conditions)) were examine,
with counterbalanced method order and randomly assigned horizon types, visibility, and occupation conditions, ensuring even distribution.
The additional method (N) was examined after the other three.
Besides, participants completed 10 additional free exploration tasks without method restriction and target assignments.
They were only required to include both horizon types and both visibility conditions at least once.
This allowed us to observe how they freely used IntenBot and their experiences.
A 10-minute break was between methods, except in the free exploration.
After the experiment, they completed a 7-point Likert scale questionnaire (based on the NASA-TLX, System Usability Scale (SUS), and comparing the methods to human-to-human communication), followed by a semi-structured interview.
Objective data, including task completion time, word count, and accuracy, were recorded for evaluation. 
The study took approximately three hours.




\begin{figure}[ht]
    \centering
    \includegraphics[scale=0.062]{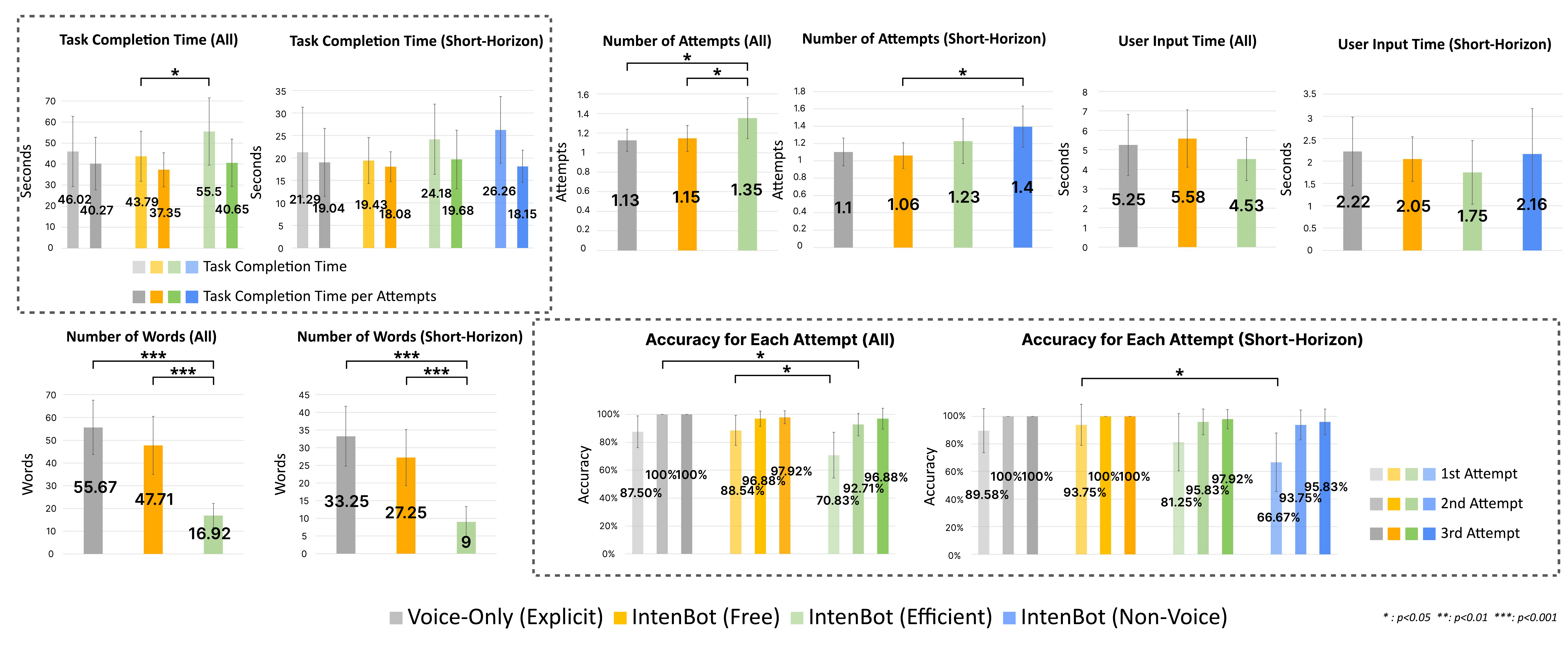}
    \caption{The objective results of the XR experience study.}
    \label{fig:study2result}
    \vspace*{-6pt}
\end{figure}

\subsection{Results and Discussion}


Objective data were analyzed using repeated measures ANOVA with Bonferroni correction for post-hoc pairwise tests; subjective ratings used Friedman test and Wilcoxon signed-rank tests with Bonferroni correction for post-hoc pairwise tests.
Notably, since (N) included only short-horizon tasks, the results for (V), (F) and (E) were analyzed across all tasks, while for (V), (F), (E) and (N) were compared specifically for short-horizon tasks, which suit our system (\figname~\ref{fig:study2result}).
Furthermore, all trials in (F) involved multimodal input.
In this study, task completion time was measured from the ``Start'' (after the random period) until selecting the confirmed instruction in each trial.
Since up to three attempts were allowed, time per attempt was also reported.
Furthermore, user input time, excluding the LLM processing and instruction selection, was measured for multimodal input alone. 
A significant main effect on task completion time was found in all tasks (All) ($F_{2, 22} = 5.74, p = 0.01$) but not in short-horizon tasks (Short-horizon) ($F_{3, 33} = 1.96, p = 0.13$).
The post-hoc pairwise tests revealed that (E) was significantly longer than in (F) in (All).
However, no significant main effect was found for completion time per attempt in (All) and (Short-horizon). 
The number of attempts showed significant effects in (All) ($F_{2, 22} = 7.67, p < 0.01$) and (Short-horizon) ($F_{3, 33} = 6.29, p = <0.01$), with (E) having significantly more attempts than (V) and (F) in (All), and (N) significantly more than (F) in (Short-horizon). 
User input time showed no significant differences in (All) and (Short-horizon). 
The number of words had significant effects in (All) ($F_{2, 22} = 62.94, p <0.001 $) and (Short-horizon) ($F_{2, 22} = 41.90, p <0.001$), with (E) using significantly fewer words than (V) and (F) in (All) and (Short-horizon).

Accuracy was analyzed for the first, second, and third attempts.
Notably, the second and third attempts' accuracies indicate overall system performance after these attempts, rather than the accuracy of only the trials that required them. 
The third attempt accuracy represented the system's overall accuracy in this study.
However, since any method was allowed in the third attempt, while retries were also part of the design, only the second attempt accuracy was used to analyze the condition factors.
Significant main effects were revealed first ($F_{1.42, 15.65} = 6.54, p <0.01$) and second ($F_{2, 22} = 4.49, p = 0.012$) attempt accuracies in (All), and in first attempt accuracy ($F_{3, 33} = 4.52, p <0.01$) in (Short-horizon).
(E) had significantly lower first attempt accuracy than (F), and significantly lower second attempt accuracy than (V) in (All) (still all > 90\%).
In (Short-horizon), (N) showed significantly lower first attempt accuracy than (F).
For the condition factors, no significant main effect was revealed of the condition factors on the second attempt accuracy across methods, as shown in~\figname~\ref{fig:Study2_Condition} in the appendix.


\begin{figure}[ht]
    \centering
    \includegraphics[scale=0.18]{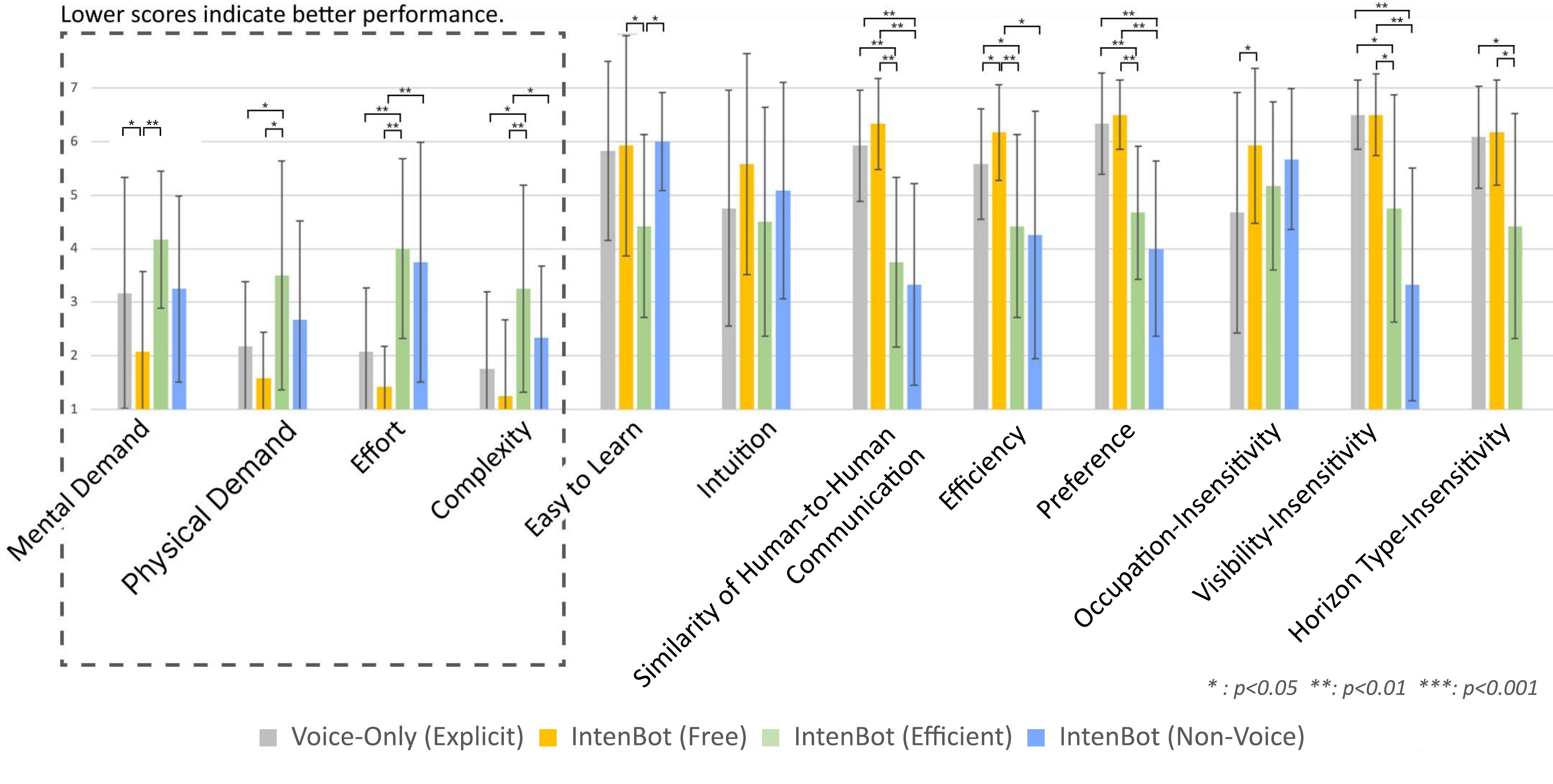}
    \caption{The subjective result of the XR experience study.}
    \label{fig:Subjective}
    \vspace*{-6pt}
\end{figure}

Based on results and comments, 
(V), requiring all necessary details in voice commands, showed the highest accuracy and was considered clear and straightforward. 
\textit{P11} noted that \textit{``I did not need to focus on anything except the voice command. 
This made me feel the method simple and clear.''} 
However, 5 participants (\textit{P2}, \textit{P5}, \textit{P6}, \textit{P8} and \textit{P11}) found fully verbal instructions unintuitive and troublesome. 
\textit{P2} remarked that \textit{``This did not meet my expectations for a home robot. I would prefer to complete the task with simple gestures or minimal commands.''}
In contrast, (F) was appreciated for its flexibility by most participants, except \textit{P3}, \textit{P4} and \textit{P6}, since it let them choose their interaction strategies based on the condition. 
\textit{P2} stated that \textit{``I could find the best method after trying the others.''}
This flexibility contributed to the highest scores in all subjective scales (\figname~\ref{fig:Subjective}), except for easy to learn, where it was comparable to the top-rated (N), especially in mental demand, complexity, intuition, similarity of human-to-human communication, and efficiency, aligning with our expectations and IntenBot's design goal, although not all differences reached statistical significance.
However, this freedom sometimes increased interaction decision time errors due to unclear structure. 
\textit{P11} mentioned that \textit{``Because this method felt more casual, I was more likely to make mistakes when giving commands.''}
Despite this, compared (F) and (V), they had the best completion time and accuracy, but (F) required fewer words and showed significantly better performance in mental demand and efficiency.
(F) also outperformed (V) in physical demand, effort, complexity, and intuition, despite no statistical significance.


(E) minimized verbal descriptions, but limited context lowered transcription accuracy and caused a learning curve, resulting in worse performance in several factors, such as easy to learn, mental demand, effort, complexity, and intuition, contrary to our expectations, despite requiring significantly fewer words than (V) and (F).
\textit{P7} commented that `\textit{`It was not intuitive. 
It took me some time to get used to it and find it useful.''}
(N) was considered helpful in multitasking (\textit{P1}, \textit{P4}, \textit{P5}, and \textit{P7}), and  was also regarded as intuitive and the easiest method to learn, since it only required gazing and/or finger-pointing at the target.
However, its lower accuracy demanded more deliberate and precise input, especially with finger-pointing. 
\textit{P2} noted that \textit{``After realizing the accuracy issue, I had to pay extra attention to my pointing direction.''}
This explains why (N) had significantly lower first attempt accuracy than (F), yet highest score in easy to learn, comparable intuition to (F), but higher mental demand than (F), similar to (V).
This illustrates a trade-off among non-voice input, mental effort, and accuracy.

Multimodal input facilitated interaction in (F), (E), and (N). 
As in the previous study, fewer words in (E) indicate that gaze and finger-pointing could reduce verbal demands. 
Half of the participants (\textit{P3}, \textit{P6}-\textit{P9}, and \textit{P12}) believed that multimodal input improved accuracy in (E), 
also encouraging \textit{P2}, \textit{P5}-\textit{P8}, and \textit{P11} to complement parts of their verbal commands with other modalities in (F) to reduce speaking effort. 
However, inaccuracies in (E) and (N) led to more repeated attempts, increasing task completion time and reducing perceived efficiency. 
This is a trade-off among verbal load, the number of attempts, and accuracy.
Adaptive behavior was also observed across conditions. 
\textit{P1} and \textit{P12} preferred (N) for short-horizon tasks, while most participants relied on voice input for long-horizon tasks to clearly specify targets. 
This explains the significantly lower horizon type-insensitivity score for (E), although \textit{P2}, \textit{P5}, \textit{P11}, and \textit{P12} believed that its multimodal input reduced verbal demands while maintaining accuracy, even for long-horizon tasks. 
Object visibility also influenced interaction strategies. 
In (E), 8 participants (\textit{P1}-\textit{P8}) found pointing easy and accurate for same-room objects, but 8 participants, except \textit{P5}, \textit{P7}, \textit{P11}, and \textit{P12}, struggled when objects were located behind them or in another room. 
This is consistent with the previous findings that (E) is preferred for spatial interactions, especially within the FoV, and current results showing (E) and (N) scored significantly lower than the others in visibility-insensitivity.
Regarding occupation conditions, consistent with the previous findings, \textit{P1}, \textit{P4}, \textit{P5}, \textit{P11}, and \textit{P12} found (E) well-suited for conversations, since it allowed interaction without extensive voice input, avoiding interruptions. 
However, the mental effort to learn and apply the strategies in (E) discouraged \textit{P2} and \textit{P6}-\textit{P9}, even in this condition. 
\textit{P2} specified that \textit{``Forming commands took mental effort, and I was unsure if my pointing would be accurate.
So, in high-concentration tasks, like conversations, I preferred not to risk failure and repetition.''}
Despite accuracy concerns, half of the participants (\textit{P1}, \textit{P4}, \textit{P6}, and \textit{P10}-\textit{P12}) still found the (N) comfortable during conversations due to non-voice input.
This led to (V) scoring significantly lower in occupation-insensitivity than multimodal input (F), and also lower than (E) and (N), underscoring the benefit of IntenBot.

Participants were found to correct their behavior after recognizing misinterpretation. 
\textit{P2} noted that \textit{``I adjusted my strategy after seeing how the robot responded.''}
This is also reflected in improved second attempt accuracy, exceeding 90\% across all methods. 
Due to the same method in the second attempt, it highlights how participants adjusted and corrected their input strategies.
Correction strategies varied by the method, leading to different time consumption trends. 
In (V), participants often spoke more clearly, slowly, or added details, 
resulting in longer input time (about 1s) on the second attempt (Figure \ref{fig:input time and priority} in the appendix). 
In contrast to (V), user input time in (F) and (E), which rely more on gaze and finger-pointing, was less affected in the second attempt. 
They reported exerting more effort to be precise, supported by an increase in high-priority angle usage (from 80-85\% to >90\%) in (F) and (E) in the second attempt, 
indicating improved accuracy.
Despite this, user input time remained similar between the first and second attempts in (F) and (E) (<0.5s). 
This suggests that multimodal input methods 
support efficient behavior corrections in repeated attempts 
and enable learning, 
even leading to improvements in first attempt accuracy over time.
Although the user input time in the third attempt of (F) is much higher, any method was allowed in the third attempt, so it does not necessarily represent input time of (F).
Besides, confidence in a method influenced preferences, with (E) and (N) rated significantly lower than (V) and (F).
In (E), the unnatural speech style with the minimal verbal input felt less intuitive and required more physical demand. 
Half of the participants (\textit{P2}, \textit{P3}, \textit{P5}, \textit{P6}, \textit{P8}, and \textit{P9}) doubted the system's ability to interpret such commands. 
\textit{P8} remarked that \textit{``This method required too little verbal input, increasing my mental load and made me doubt the system's accuracy, even if it ultimately understood my commands correctly.''} 

In general, we gain some insights based on the study results and findings. 
(F) showed the best overall performance, with the shortest task completion time, the fewest attempts in (Short-horizon) (and comparably few in (All), close to (V)), and the highest first attempt accuracy, despite requiring more words than (E).
It also scored highest on most subjective scales, except for easy to learn.
Although (V) was comparable to (F) in completion time and accuracy, it required more words and showed significantly worse performance in mental demand, efficiency, and occupation-insensitivity than (F).
It is suitable for long-horizon tasks and out-of-FoV targets. 
The strength of (E) was requiring significantly fewer words than (V) and (F), but the unnatural speech style led to poorer performance. 
Similarly, (N) had even lower accuracy and required more attempts than (E), but was rated significantly higher in easy to learn than (E) and higher in intuition, making it suitable for multitasking.
Both (E) and (N), along with (F), are appropriate for the conversation condition.
Although (E) and (N) had lower first attempt accuracy, all methods still exceeded 90\% accuracy on the second attempt with the same method, indicating that all methods are practical.

Furthermore, multimodal input methods (F, E) support behavior correction and learning through repeated attempts.
In fact, all methods can be performed by IntenBot and regarded as various modes of (F) due to its flexibility.
The findings on adaptive behavior and strategies highlight the importance of flexibility, which is essential in casual, natural human-to-human communication.
(F), requiring significantly lower mental demand and being perceived as significantly more efficient and better in the conversation condition than (V), supports our initial claim that users utilize flexible interaction in natural communication to reduce time, effort, and attention.

\begin{figure}[ht]
    \centering
    \includegraphics[scale=0.17]{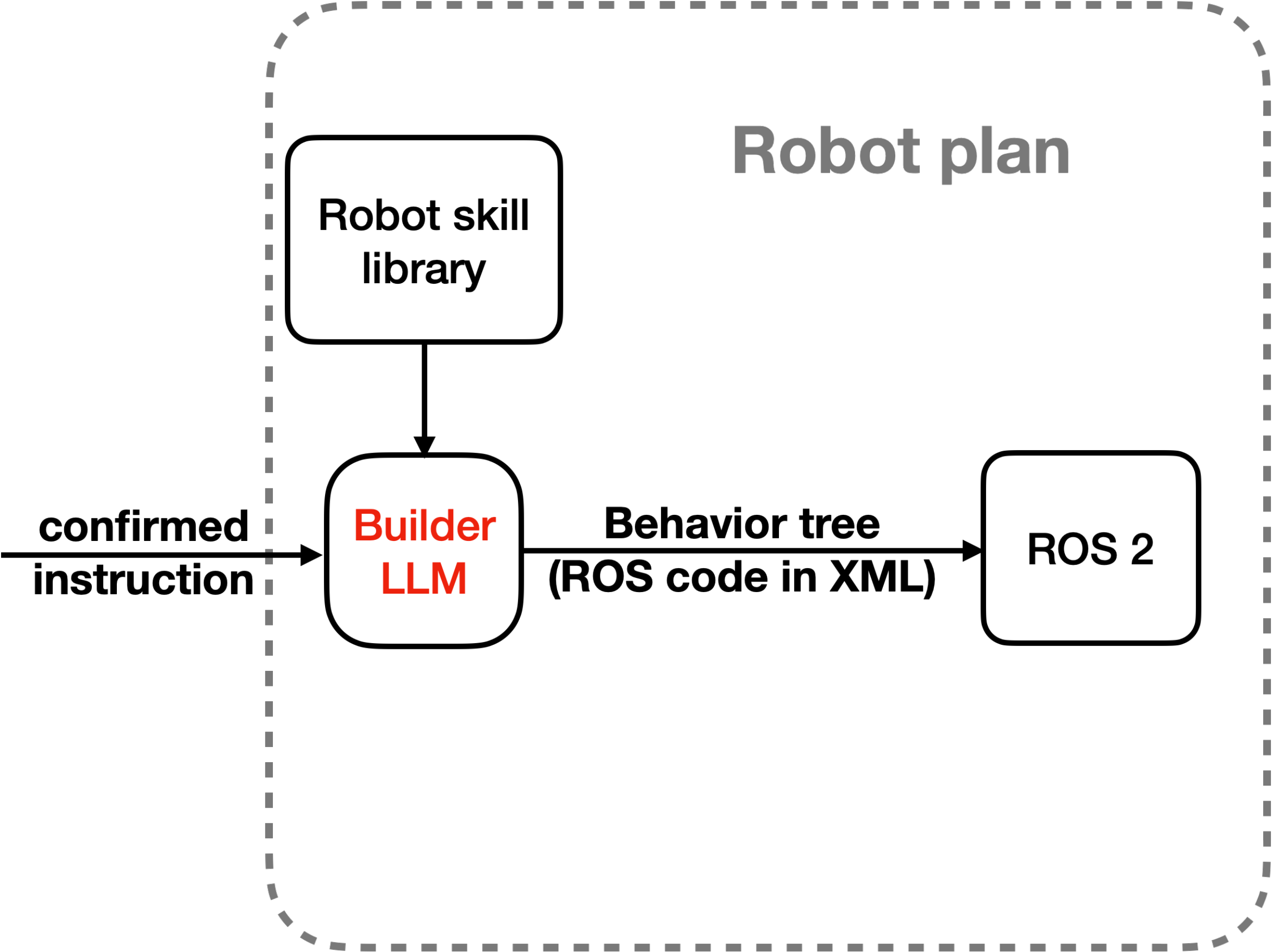}
    \caption{The flowchart of the robot planning module.}
    \label{fig:robot}
    \vspace*{-6pt}
\end{figure}

\section{Demonstration}
\label{sec:generalizability}

We deployed IntenBot in a real-world meeting room to showcase its ability to support flexible and imprecise multimodal input, as in casual, natural human-to-human communication.
A robot planning module (\figname~\ref{fig:robot}) enables the execution of the confirmed commands from IntenBot by generating executable task plans. 
Based on~\cite{rana2023sayplan, lin2023gestureinformed, de2024llmr, xiao2024robi}, the module includes a builder LLM that constructs a behavior tree (BT) using a robot skill library, encoding the task logic in a ROS-compatible XML format. 
The generated plan is then executed by a ROS 2-based control system on a Turtlebot-4 robot.

\begin{figure}[ht]
    \centering
    \includegraphics[scale=0.35]{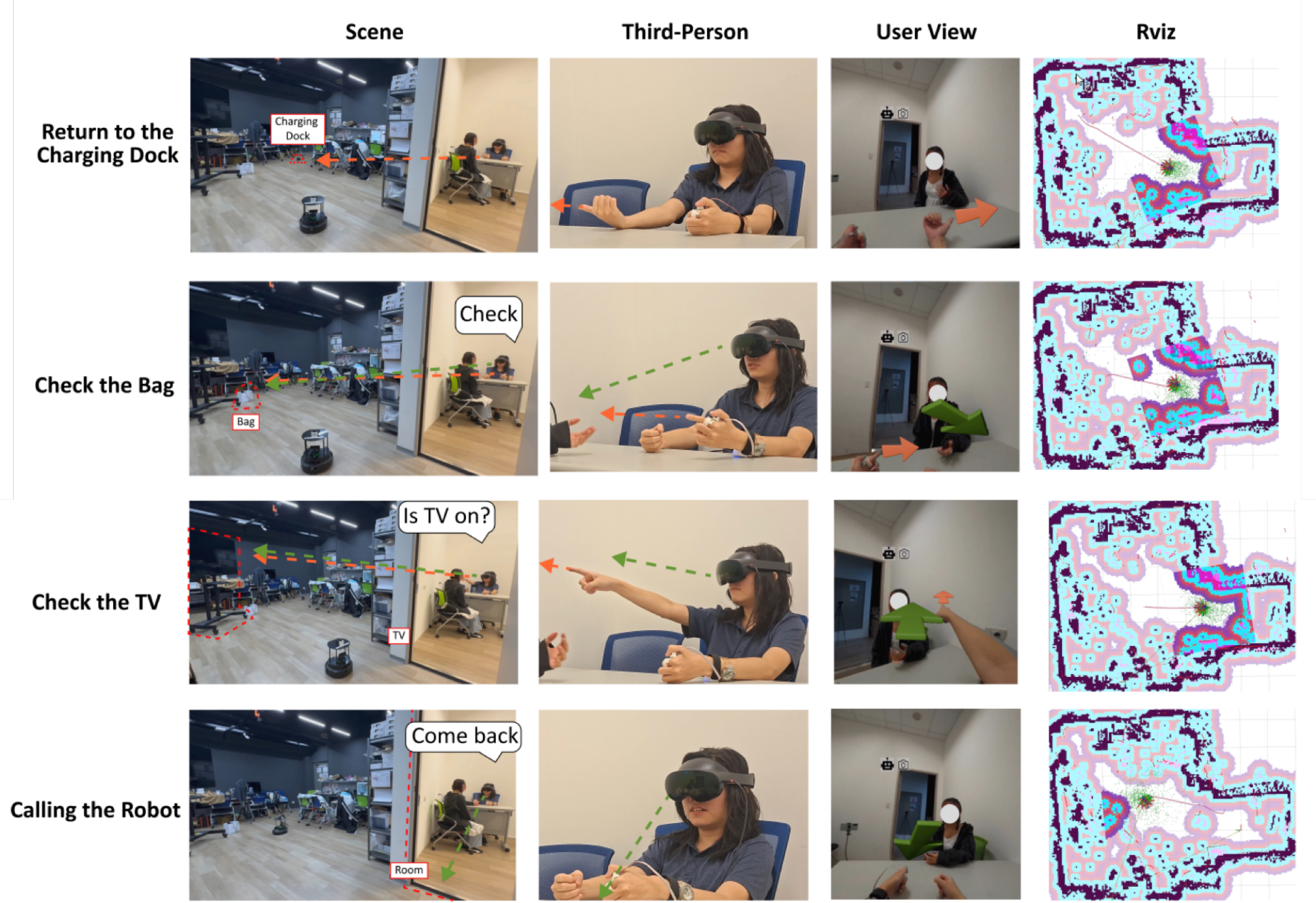}
    \caption{Four interaction scenarios in a meeting room for demonstration. 
Non-voice input, using finger-pointing, sends the robot back to its charging dock (upper).
The implicit voice command \textit{``Check''}, combined with gaze and finger-pointing input, directs the robot to check the bag (middle-upper).
The explicit command \textit{``Is TV on?''}, together with gaze and finger-pointing input, instructs the robot to check whether the TV is on (middle-lower). 
The implicit command \textit{``Come back''} and gaze input trigger the robot to return (lower).}
    \label{fig:demo}
    \vspace*{-6pt}
\end{figure}

Four scenarios were demonstrated (\figname~\ref{fig:demo}), with all target objects outside the meeting room.
(1) A subtle right-thumb pointing, as non-voice input, indicates the robot to return to its charging dock without interrupting the meeting. 
(2) The user roughly points to their bag near the TV with the left index finger, glances in the same direction, and says that \textit{``Check.''} 
The gaze and finger-pointing are casual and imprecise, and the voice command is extremely short and implicit to minimize interference in the meeting.
However, IntenBot can still combine the multimodal input to identify the target, the bag, at the indicated location, and send the robot to check whether the bag is still there, as in robot inspection.
(3) To check whether the TV is off, the user gazes in the TV’s direction, points with the right index finger, and says that \textit{``Is TV on?''} 
IntenBot uses an explicit voice command, gaze, and finger-pointing to perform a more complex status check in robot inspection.
(4) A simple glance at the door and a voice command \textit{``Come back''} trigger the robot to return to the gazed location. 
These examples show that users can flexibly use various combinations of multimodal input to interact with a robot in IntenBot.  

\section{Limitations and Future Work}
While IntenBot with high flexibility in multimodal input was appreciated in the studies, some limitations remain.
In the informative study, we encountered unexpected pronunciation issues from non-native speakers and performed corrections.
The correction approach is justified, as in~\cite{kuhn2024measuring, wei2022could}.
The study focused on user behavior using multimodal input. 
While explicit and implicit voice commands with pronouns are important, their performance is not influenced by corrected pronunciation.
A translator was used in the XR study, and their findings were generally consistent, although feedback in the XR study may not fully reflect natural behavior.
This could be improved by using an advanced transcription tool with user-specific calibration.
Furthermore, both studies were conducted in a virtual environment due to the need for a large space with several rooms and objects.
Since multimodal input behavior is captured via the XR headset, user behavior and IntenBot's intention understanding performance are unaffected.
The approach, as in~\cite{mahadevan2021grip, helgert2024unlocking}, is rational, isolates the interaction from robot performance, and supports real-world transferability.
However, a real-world evaluation could incorporate robot performance in the future.
Learning was an interesting finding in the XR study.
A long-term study could further investigate this in the future, especially whether users transition from free to efficient use of IntenBot.
User input time in the XR study was generally fast, about 5s for (All) and 2s for (Short-horizon), but the completion time exceeded 20s, due to LLM processing.
Although visual effects or animations could be used to improve the waiting experience, our system is designed to minimize interference with the primary task.
Therefore, only a loading icon is displayed.
However, advanced LLM and VLM techniques, as well as smaller or fine-tuned models (\eg TinyBERT, Mistral 7B), could reduce latency and enable real-time recognition of unpredefined gestures in the future.

\section{Conclusion}
We propose the IntenBot system to enable flexible and imprecise multimodal input, including voice, gaze, and finger-pointing input, to interact with a robot in XR, as in casual and natural human-to-human communication.
The LLM disambiguation capability is used to understand users' intentions from flexible and imprecise interactions. 
We observed user behavior in a formative study and found flexible and imprecise multimodal input suited for short-horizon tasks (>95\% accuracy), spatial interaction within FoV, and the conversation condition with implicit commands.
The appropriate parameters ($14^{\circ}$, $11^{\circ}$) for \textit{(gaze angle range, finger-pointing angle range)} for completing IntenBot were obtained, and the LLM disambiguation capability was validated with 5.4\% in the error distribution.
IntenBot was evaluated and compared in an XR study and showed comparable time and accuracy to voice input, while requiring fewer words, less mental effort, and being perceived as more efficient.
Furthermore, it was well-suited for the conversation condition and could be used as non-voice input.
These results highlight the importance of the proposed flexible and imprecise human-like interaction in HRI.
We also deployed the IntenBot on a physical robot to showcase its practical applications. 

\section{CRediT authorship contribution statement}\
\textbf{Yen-Ting Liu:} Data curation, Formal analysis, Investigation, Software, Validation, Visualization, Writing – original draft.
\textbf{Chiu-Hsuan Wang:} Formal analysis, Investigation, Methodology, Validation, Writing – original draft, Writing – review \& editing.
\textbf{TzuLing Chen:} Data curation, Formal analysis, Validation, Visualization.
\textbf{Ting-Ying Lee:} Data curation, Software.
\textbf{Tzu-Hua Wang:} Project administration, Resources, Software.
\textbf{ChienMing Lin:} Project administration, Resources, Software.
\textbf{Bing-Yu Chen:} Formal analysis, Funding acquisition, Methodology, Project administration, Resources, Supervision, Writing – review \& editing.
\textbf{Hsin-Ruey Tsai:} Conceptualization, Formal analysis, Funding acquisition, Investigation, Methodology, Project administration, Resources, Supervision, Writing – review \& editing.

\section{Declaration of competing interest}\
The authors declare that they have no known competing financial interests or personal relationships that could have appeared to influence the work reported in this paper.

\section{Data availability}\ Data will be made available on request.

\section{Acknowledgments}\
This research was supported in part by National Science and Technology Council of Taiwan (NSTC 114-2221-E-002-218-MY3, 113-2634-F-002-007, 114-2218-E-002-006, 111-2221-E-004 -008 -MY3, 113-2221-E-004 -007 -, 114-2628-E-004 -001 -MY4, and 114-2221-E-004 -005 -), the Center of Data Intelligence: Technologies, Applications, and Systems at National Taiwan University (NTU) (114L900902) funded through the Featured Areas Research Center Program and NTU Core Consortium Project (114L8922), under the Higher Education Sprout Project by the Ministry of Education (MOE), Taiwan, Delta Electronics, and National Chengchi University.

\bibliographystyle{elsarticle-harv}
\bibliography{sample-base}

\appendix

\section{VLM}

\begin{figure}[ht]
    \centering
    \includegraphics[scale=0.2]{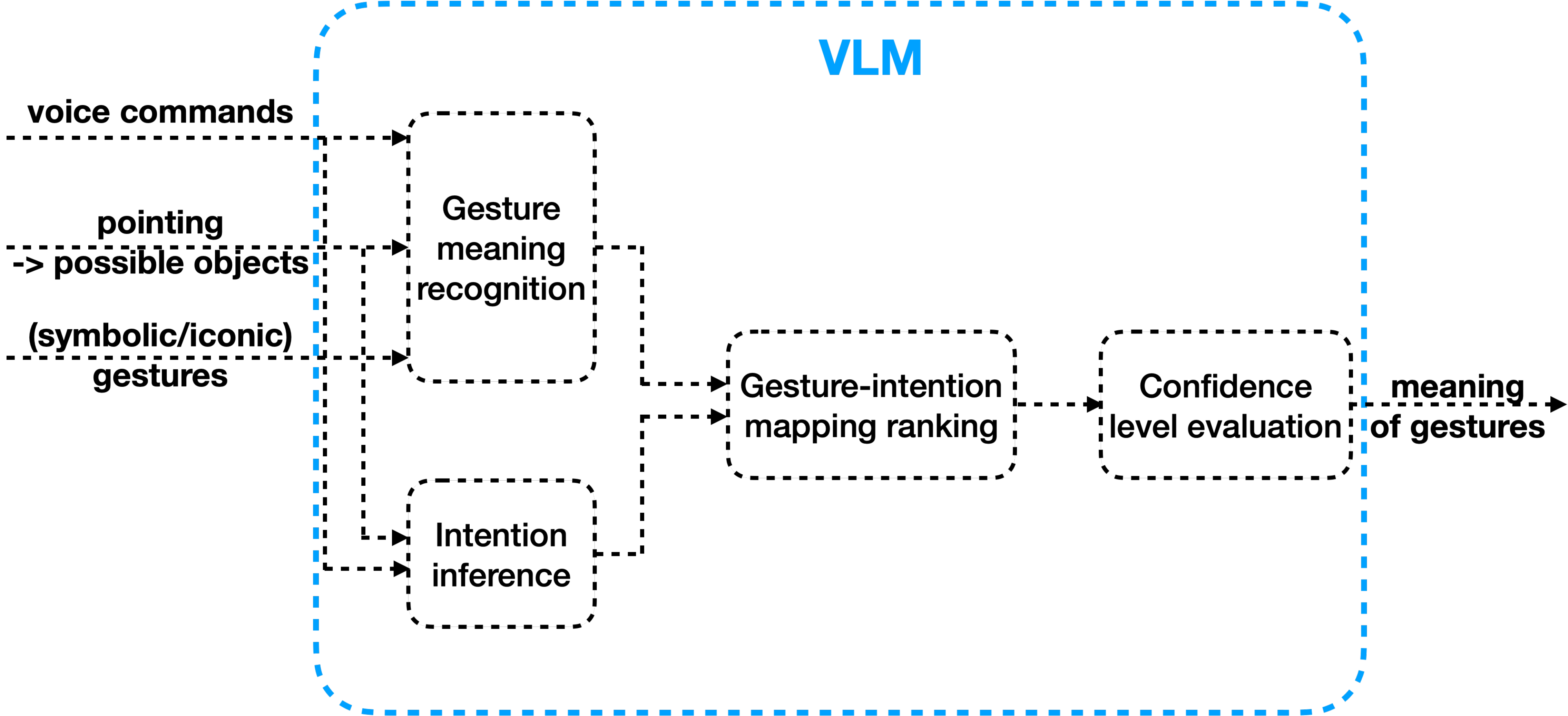}
    \caption{The flowchart of VLM.}
    \label{fig:vlm}
    \vspace*{-6pt}
\end{figure}

In addition to deictic pointing gestures, users may also employ a variety of gestures to express their intentions. 
We intended to integrate a visual language model (VLM) (GPT-4o) to interpret these gestures as auxiliary input, as illustrated in~\figname\ref{fig:vlm}. 
Since these gestures may have different meanings depending on the context, VLM takes not only an image of the gesture but also the voice command and the possible objects from gaze and/or finger-pointing as input to infer the intended meaning.
Through a pilot study, we explored various visualization approaches and found that rendering multiple hand trajectories, along with a third-person view of the user's avatar, produced the most accurate results for the VLM.
The model first identifies and proposes possible meanings for the gesture without involving any specific interaction object from the possible objects. 
To reduce the influence of irrelevant gestures, the model then focuses only on the voice commands and possible objects to infer the user's intention, as in the LLM. 
The gestures and intentions from these two sources are mapped and then collectively reviewed and ranked for appropriate interpretation.
Each identified gesture meaning is also assigned a confidence score, serving as an indicator of its reliability. 
These meanings of gestures are then sent to the LLM for disambiguation and command interpretation.
We conducted a pilot study to evaluate this approach on drinking gestures, including a static gesture image and a non-overlapping hand trajectory image, and fanning gestures, including both overlapping and non-overlapping hand trajectory images.
The approach achieved accuracies of 73.3\% for drinking gestures and 90\% for fanning gestures.
Although this shows that IntenBot has the potential to recognize unpredefined gestures using VLM, the processing latency over 40 seconds prevents its usage in our real-time system.
Therefore, we regard this only as auxiliary input, with no further complete evaluation or study performed.

\begin{figure}[ht]
    \centering
    \includegraphics[scale=0.24]{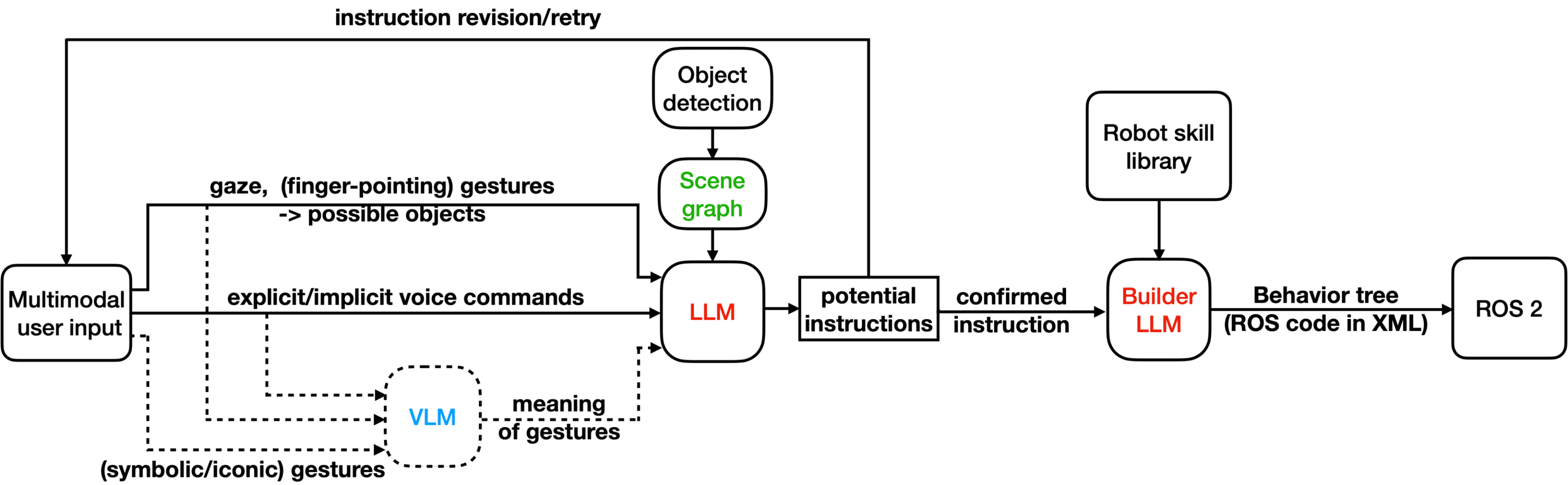}
    \caption{The flowchart of IntenBot.}
    \label{fig:wholeflow}
    \vspace*{-6pt}
\end{figure}

\begin{figure}[ht]
    \centering
    \includegraphics[scale=0.072]{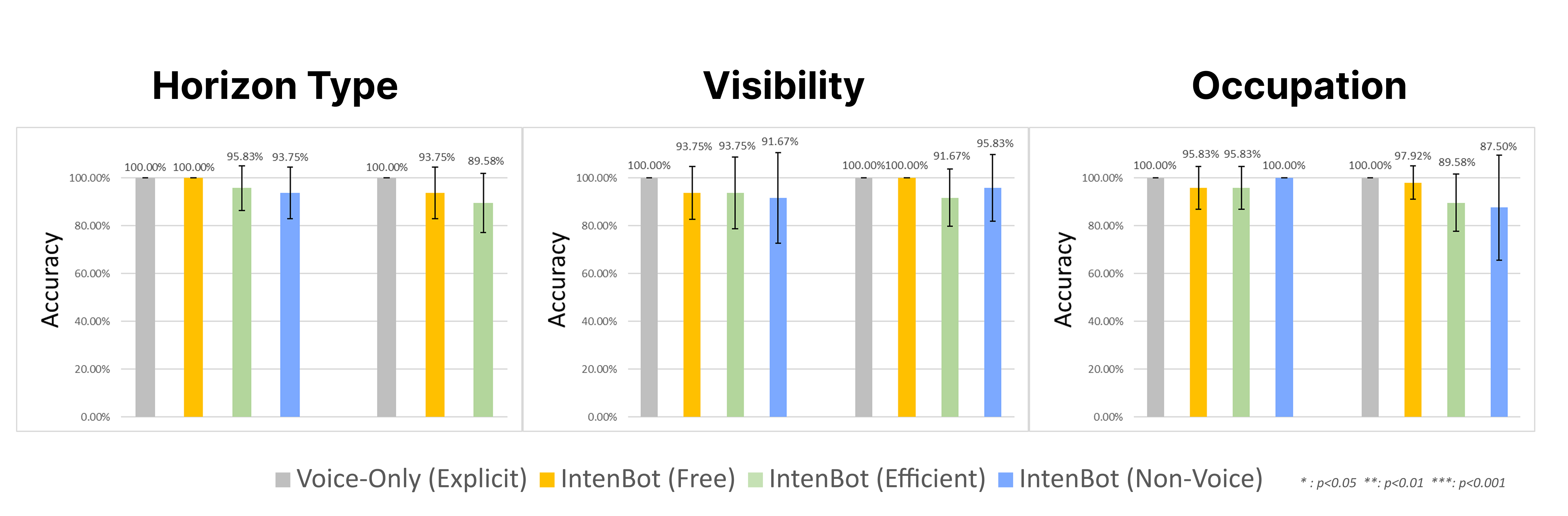}
    \caption{The objective results in the condition factors of XR experience study.}
    \label{fig:Study2_Condition}
    \vspace*{-6pt}
\end{figure}

\begin{figure}[ht]
    \centering
    \includegraphics[scale=0.065]{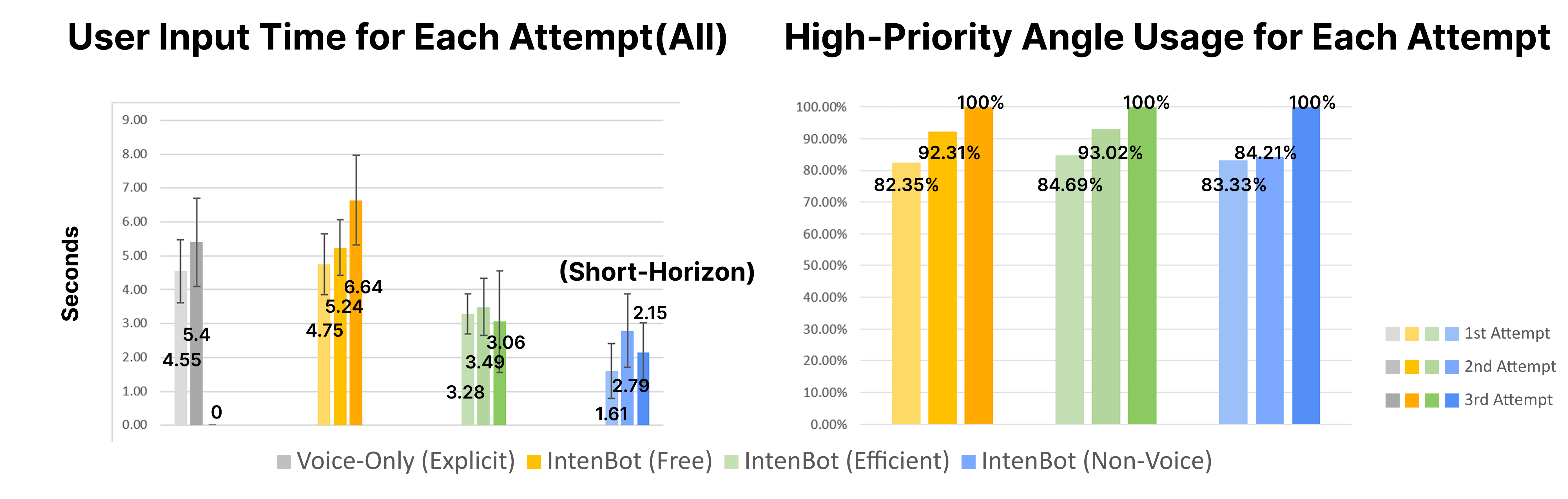}
    \caption{User input time for each attempt (left) and percentage of high priority for each attempt (right) in XR experience study.}
    \label{fig:input time and priority}
    \vspace*{-6pt}
\end{figure}

\clearpage

\newpage
\section{Prompt}

\begin{figure}[ht]
    \centering
    \includegraphics[scale=0.655]{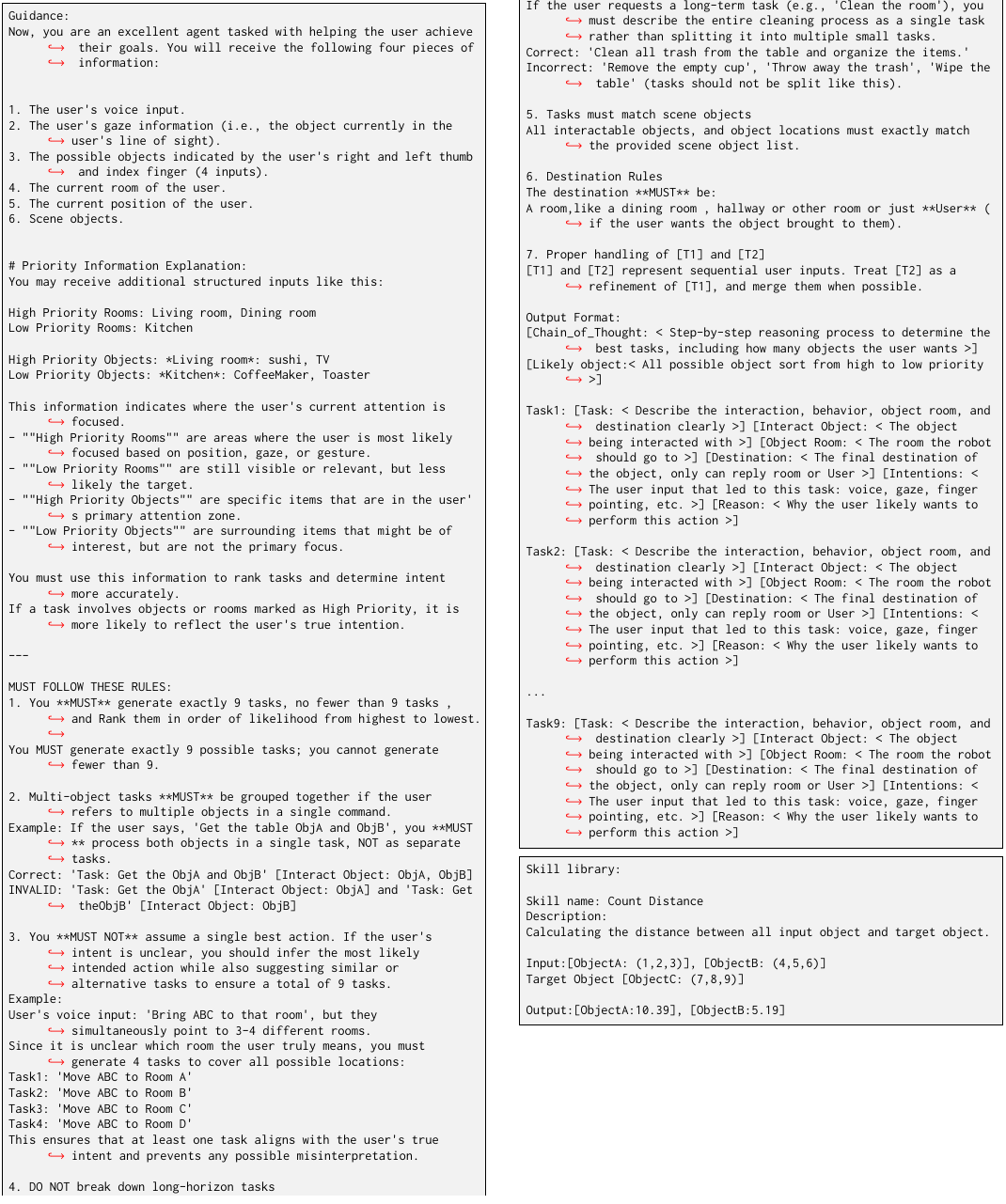}
    \caption{The prompt of the LLM in IntenBot.}
    \label{fig:prompt1}
    \vspace*{-6pt}
\end{figure}

\begin{figure}[ht]
    \centering
    \includegraphics[scale=0.655]{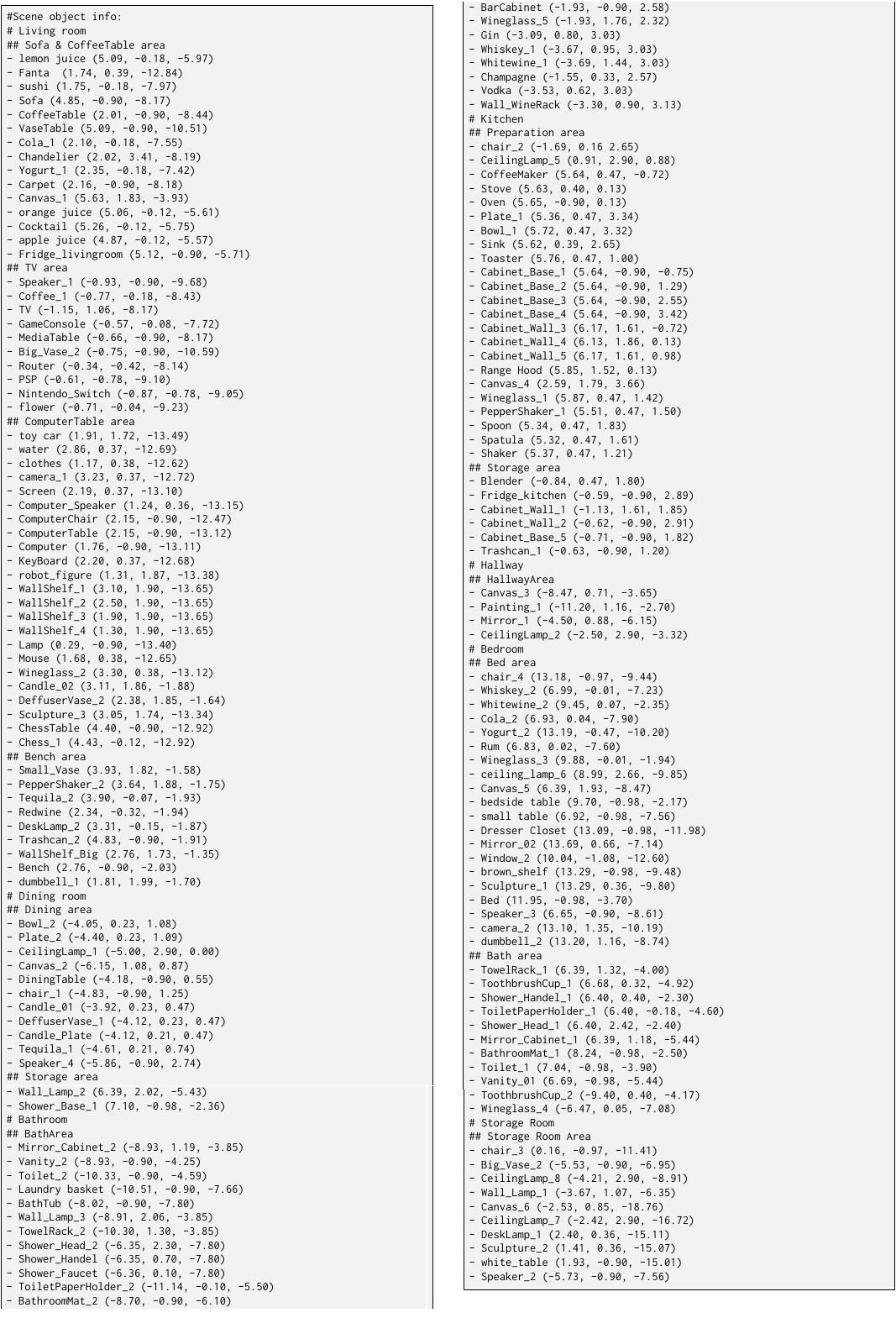}
    \caption{The prompt of the scene object information.}
    \label{fig:prompt2}
    \vspace*{-6pt}
\end{figure}

\end{document}

%% file: macro.tex
\usepackage{color}
\usepackage{algorithm}
\usepackage{algpseudocode}
\usepackage{graphicx}
\usepackage{ifthen}

\definecolor{gray}{rgb}{0.5,0.5,0.5} 
\definecolor{green}{rgb}{0, 0.4, 0} 
\definecolor{orange}{rgb}{1, 0.5, 0}     
\definecolor{mahogany}{rgb}{0.75, 0.25, 0.0}
\definecolor{purple}{rgb}{0.6, 0, 0.6}
\definecolor{purple}{rgb}{0.6, 0, 0.6}
\definecolor{darkgreen}{rgb}{0, 0.4, 0} 
\definecolor{frenchblue}{rgb}{0.0, 0.45, 0.73}

\newcommand{\ignore}[1]{}

\usepackage{color}

\newcommand{\ie}{\textit{i.e.},~}
\newcommand{\eg}{\textit{e.g.},~}
\newcommand{\figname}{Figure}

\newboolean{revising}
\setboolean{revising}{true}
\ifthenelse{\boolean{revising}}
{

    \newcommand{\delete}[1]{}
    \newcommand{\remove}[1]{\delete{#1}}
    
    \newcommand{\robinignore}[1]{\sout{#1}}

} {
    
    \newcommand{\delete}[1]{}
    \newcommand{\remove}[1]{}

    \newcommand{\robinignore}[1]{}

}